\documentclass[11pt,letterpaper]{article} 
\usepackage{cite}
\usepackage{algorithmic}
\usepackage{graphicx}
\usepackage{textcomp}
\usepackage{xcolor}
\usepackage{amsmath,latexsym,amssymb,color,amsthm,amsfonts}
\usepackage{ifthen,graphics,epsfig}
\usepackage{xspace}
\usepackage{enumitem}
\usepackage{url}

\usepackage{float}
\newfloat{algorithm}{thp}{lop}
\floatname{algorithm}{Algorithm}
\newcommand{\Xomit}[1]{}

\usepackage{marginnote}
\usepackage{todonotes}

\newcommand{\remove}[1]{}

\newlength {\squarewidth}

\newtheorem{theorem}{Theorem}
\newtheorem{lemma}{Lemma}

\newcommand{\toto}{xxx}
\newenvironment{proofT}{\noindent{\bf
Proof }} {\hspace*{\fill}$\Box_{Theorem~\ref{\toto}}$\par\vspace{3mm}}
\newenvironment{proofL}{\noindent{\bf
Proof }} {\hspace*{\fill}$\Box_{Lemma~\ref{\toto}}$\par\vspace{3mm}}

\newcounter{linecounter}
\newcommand{\linenumbering}{\ifthenelse{\value{linecounter}<10}{(0\arabic{linecounter})}{(\arabic{linecounter})}}
\renewcommand{\line}[1]{\refstepcounter{linecounter}\label{#1}\linenumbering}
\newcommand{\resetline}[1]{\setcounter{linecounter}{0}#1}
\renewcommand{\thelinecounter}{\ifnum \value{linecounter} > 9\else 0\fi \arabic{linecounter}}

\newcommand{\coordval}{{\sc{coord\_value}}}
\newcommand{\coord}{\mathit{coord}}
\newcommand{\send}{\mathit{\sf send}}
\newcommand{\sto}{\mathit{\sf to}}
\newcommand{\receive}{\mathit{\sf receive}}
\newcommand{\broadcast}{\mathit{\sf broadcast}}
\newcommand{\valid}{\mathit{\sf valid}}
\newcommand{\ttrue}{\mathit{\tt true}}

\newcommand{\proposals}{\mathit{proposals}}
\newcommand{\bindec}{\mathit{bin\_decisions}}
\newcommand{\BINCONS}{\mathit{BIN\_CONS}}
\newcommand{\binpropose}{{\sf bin\_propose}}

\newcommand{\mvpropose}{{\sf mv\_propose}}
\newcommand{\RBbroadcast}{{\sf RB\_broadcast}}
\newcommand{\RBdeliver}{{\sf RB\_deliver}}
\newcommand{\wait}{{\sf wait\_until}}

\newcommand{\mmin}{\mathit{\sf min}}
\newcommand{\BVbroadcast}{{\sf BV\_broadcast}}
\newcommand{\BAMP}{{\cal BAMP}_{n,t}}
\newcommand{\ES}{\mathit{\Diamond Synch}}
\newcommand{\binvalues}{\mathit{bin\_values}}
\newcommand{\values}{\mathit{values}}
\newcommand{\set}{{\sf set}}
\newcommand{\modulo}{{\sf ~mod~}}
\newcommand{\decide}{{\sf decide}}

\newcommand{\firstzero}{{\mathit{first}_0}}
\newcommand{\first}{{\mathit{first}}}
\newcommand{\firstr}{{\mathit{first}_r}}
   
\newcommand{\last}{{\mathit{last}}}
\newcommand{\lastr}{{\mathit{last}_r}}
\newcommand{\lastzero}{{\mathit{last}_0}}

\newcommand{\lastrho}{{\mathit{last}_\rho}}
\newcommand{\firstrho}{{\mathit{first}_\rho}}

\newenvironment{smallenum}{
\begin{itemize}[leftmargin=*]
    \setlength{\topsep}{-3pt} 
    \setlength{\partopsep}{-3pt}
  \setlength{\itemsep}{0pt}
  \setlength{\parskip}{-1pt}
  \setlength{\parsep}{-6pt}
}{\end{itemize}}

\begin{document}

\title{DBFT: Efficient Byzantine Consensus with a Weak Coordinator and its Application to Consortium Blockchains
\footnote{%
Tyler Crain and Vincent Gramoli were supported by the Australian
Research Council's Discovery Projects funding scheme (project number
160104801).  Vincent Gramoli is the recipient of the Australian
Research Council Discovery International Award.  Mikel Larrea was
supported by the Spanish Research Council, grant TIN2016-79897-P, and
the Basque Country Research Council, grants IT980-16 and MV\_2016\_1\_0031.
Michel Raynal was supported by the French ANR project DESCARTES (grant
16-CE40-0023-03) devoted to distributed software engineering.}}

\author{Tyler Crain$^{\dag}$~~
       Vincent Gramoli$^{\dag,\ddag}$~~ 
       Mikel Larrea$^{\dag,\S}$~~
       Michel Raynal$^{\star,\circ}$\\~\\
$^{\dag}$   University of Sydney, Australia \\
{\small {\tt 
             \{tyler.crain,vincent.gramoli\}@sydney.edu.au}}\\
$^{\ddag}$   Data61-CSIRO, Australia \\
$^{\S}$  University of the Basque Country UPV/EHU, Spain\\
{\small {\tt 
             mikel.larrea@ehu.eus}}\\
$^{\star}$  Institut Universitaire de France\\
$^{\circ}$  IRISA, Universit\'e de Rennes, France \\
{\small {\tt 
             raynal@irisa.fr}}
}


\maketitle


\begin{abstract}

This paper introduces a deterministic Byzantine consensus algorithm that relies on a new \textit{weak coordinator}.
As opposed to previous algorithms that cannot terminate in the presence of a faulty or slow coordinator, our algorithm can terminate even when its coordinator is faulty, hence the name weak coordinator.
The key idea is to allow processes to complete asynchronous rounds
  as soon as they receive a threshold of messages, instead of having to wait
  for a message from a coordinator that may be slow.
%
%

The resulting algorithm assumes partial synchrony, is resilience optimal, time optimal and does not need signatures. 
Our presentation is didactic: we first present a simple safe binary Byzantine consensus algorithm, 
modify it to ensure termination, and finally present an optimized reduction from multivalue consensus to binary consensus that may terminate in 4 message delays. 


To evaluate our algorithm, we deployed it on 100 machines distributed in 5 datacenters across different continents and compared its performance against the randomized solution from Most\'efaoui, Moumem and Raynal  [PODC'14] that terminates in $O(1)$ rounds in expectation.
%
Our algorithm always outperforms the latter even in the presence of Byzantine behaviors.
Our algorithm has a subsecond average latency in most of our geo-distributed experiments, even when attacked by a well-engineered coalition of Byzantine processes.
\end{abstract}



\section{Introduction and Related Work}

To circumvent the impossibility of solving consensus in asynchronous message-passing
systems~\cite{FLP85} where processes can be faulty or \emph{Byzantine}~\cite{LSP82},
researchers typically use
randomization~\cite{A03,BE03,CR93} or additional synchrony assumptions.

Randomized algorithms can use per-process ``local'' coins or a shared ``common'' coin to solve consensus probabilistically among $n$ processes despite $t<\frac{n}{3}$  
Byzantine processes.
When based on local coins, the existing algorithms converge in $O(n^{2.5})$ expected time~\cite{KS16}.  A recent randomized algorithm without signature~\cite{MMR14} solves consensus in $O(1)$ expected time under a fair scheduler.
The fair scheduler assumption was later relaxed in an extended version~\cite{MMR15} that we refer to as \emph{Coin} in the remainder of the paper.
Unfortunately, implementing a common coin increases the message complexity of the consensus algorithm.

To avoid the need of a common coin and solve the consensus problem \emph{deterministically}, researchers have assumed partial or eventual synchrony~\cite{DLS88}.
Interestingly, these solutions typically require a unique \emph{coordinator} process, sometimes called a leader, to be non-faulty~\cite{DDS87,DLS88,CL02,MA06,KAD07,BSA14,AGZ15,LVC16}.
The advantage is that if the coordinator is non-faulty and if the messages are delivered in a timely manner in an asynchronous round, then the coordinator broadcasts its proposal to all processes and this value is decided after a constant number of message delays.
The drawback is that a faulty coordinator can dramatically impact the algorithm performance by leveraging the power it has in a round and imposing its value to all. Non-faulty processes thus have no other choices but to decide nothing in this round.

In this paper, we present a \emph{weak coordinator} alternative that does not suffer from this drawback. It allows us to introduce a new deterministic Byzantine consensus algorithm that is time optimal, 
resilience optimal and does not need signatures.
%
%
As opposed to a classic (strong) coordinator, the weak coordinator does not impose its value.
On the one hand, this allows non-faulty processes to decide a value quickly without the help of the coordinator.
On the other hand, the coordinator helps the algorithm terminating if non-faulty processes know that they proposed distinct values that might all be decided.
%
Furthermore, having a weak coordinator allows rounds to be executed optimistically without waiting for a specific message. This differs from classic  BFT algorithms~\cite{CL02}
that have to wait for a particular message from their coordinator and sometimes have to recover from a slow network or faulty coordinator.
To mitigate the problem of a slow or Byzantine coordinator, other approaches were previously explored.
Some protocols progressively reduce the time allocated to a coordinator to solve consecutive consensus instances in order to force the change of a slow 
coordinator~\cite{CWA09,ABQ13}. 
While this still requires a (strong) coordinator in each round, it favors the fastest coordinator in successive rounds.
An exponential information gathering tree was used
to terminate in $t+3$ rounds without a coordinator~\cite{BS09}.
Other solutions~\cite{DLS88,ST87} require at least $O(t)$ rounds.
By contrast our weak coordinator only helps agreement by suggesting a value
while still allowing termination in a constant number of message delays and thus differs from the classic coordinator~\cite{CT96,DLS88}
or the eventual leader that cannot be implemented in ${\BAMP}[t<n/3]$.\\

\noindent
{\bf Application to consortium blockchains.}
To motivate our algorithm, we study its applicability to the recent context of \emph{blockchains}~\cite{N08}. 
Blockchains
originally aimed at tracking ownerships of
digital assets where any Internet user could solve a cryptopuzzle before proposing, for consensus,
a block of asset transactions.
The \emph{consortium blockchains}~\cite{But16} became promising at reducing the 
amount of resources consumed by avoiding to resolve the cryptopuzzle but restricting the set of 
proposers to $n$ known processes.

These consortium blockchains seem similar to replicated state machines~\cite{L78,S90} where a sequence of commands must be decided by multiple processes. 
Some blockchains already use Byzantine fault tolerant consensus, for example, Hyperledger~\cite{ABB18} uses a consensus based on a classic coordinator~\cite{BSA14}, 
Honeybadger~\cite{MXC16} uses a randomized algorithm~\cite{MMR14} and the Red Belly Blockchain~\cite{Gra17} uses a perliminary version of the algorithm we introduce here~\cite{CGLR18}.
%
%
A slight difference with state machine replication is that the block at index
$x$ of a blockchain must embed the hash of the block decided at instance
number $(x-1)$.
This relation between instances is interesting as it entails a natural
mechanism during a consensus instance for discarding fake proposals
or, instead, extracting a \emph{valid} value out of various proposals.
%
%

We thus propose a variant of the consensus problem that allows us to extend common
definitions of Byzantine consensus, that either assume that no value
proposed only by Byzantine processes can be
decided~\cite{CFV06,MMR15,MR16}, or
that any value (i.e.,
possibly proposed by a Byzantine 
process) can be decided~\cite{DLS88,KMM03,L96,MA06,R10}.
Interestingly, the validity property we propose allows a decided value to combine multiple proposals but is less strict than interactive consistency~\cite{PSL80} or vector consensus~\cite{NCV05}: for example, it does not require the decided value to combine at least $t+1$ values proposed by correct processes. 
\\

\noindent
{\bf Geo-distributed experimentation with Byzantine coalitions.}
To validate our expectations experimentally, we deployed our consensus algorithm on 100 Amazon VMs located in 5 datacenters on different continents. We also implemented ``Coin''  the recent randomized algorithm from Moust\'{e}oui et al.~\cite{MMR14} used in the HoneyBadger blockchain~\cite{MXC16} and demonstrated that under all our workloads, our algorithm outperforms ``Coin'' that is known to terminate in $O(1)$ round in expectation. This is due to both the overhead of the coin implementation that slows down every round and the risks of being unlucky at tossing the coin by increasing the number of rounds needed to decide.

As Byzantine behaviors are known to affect drastically performance of (strong) coordinator-based consensus~\cite{CWA09,ABQ13}, 
we also implemented 4 different Byzantine attacks: Byz1 where Byzantine processes send a bit $b$ where the protocol specification expects them to send $\neg b$; Byz2 where Byzantine processes are mute; Byz3 where Byzantine processes send a combination of random and flipped values and Byz4 where Byzantine processes form a coalition to limit the progress of non-faulty nodes from one round to another by exploiting a Byzantine coordinator and sending messages without waiting. Interestingly, the 
latency exceeds slightly the second only under the Byz3 attacks.

Finally, we combine our consensus algorithm with an optimized variant of the reduction of multivalue to binary consensus of Ben-Or et al.~\cite{BKR94} to propose a novel \emph{Democratic Byzantine Fault Tolerant (DBFT)} consensus algorithm applicable to consortium blockchains that terminates in 4 messages delays in the good case, when all non-faulty processes propose the same value.
\\

\noindent
{\bf Roadmap.}
Section~\ref{sec:model} presents the model. 
Section~\ref{sec:byz-consensus} presents the binary Byzantine consensus algorithm.
Section~\ref{sec:multi-to-bin} presents the consensus definition and an application to the blockchain context and 
Section~\ref{sec:conclusion} concludes the paper.
The proofs of safety and termination as well as experimental results are deferred to the appendix.

\section{A Byzantine Computation Model}
\label{sec:model}


\noindent
{\bf Asynchronous processes.}
The system is made up of a set $\Pi$ of $n$ asynchronous sequential processes,
namely $\Pi = \{p_1,\ldots,p_n\}$; $i$ is called the ``index'' of $p_i$. 
``Asynchronous'' means that each process proceeds at its own speed,
which can vary with time and remains unknown to the other processes.
``Sequential'' means that a process executes one step at a time.
This does not prevent it from executing several threads with an appropriate
multiplexing. 
%
Both notations
$i\in Y$ and $p_i\in Y$ are used to say that $p_i$ belongs to the set $Y$.
\vspace{0.5em}

\noindent
{\bf Communication network.}
\label{sec:basic-comm-operations}
The processes communicate by exchanging messages through
an asynchronous reliable point-to-point network. ``Asynchronous''  means that
there is no bound on message transfer delays, but these delays are finite.
``Reliable'' means that the network does not lose, duplicate, modify, or
create messages. ``Point-to-point'' means that any pair of processes
is connected by a bidirectional channel. Hence, when a process receives
a message, it can identify its sender.
A process $p_i$ sends a message to a process $p_j$ by invoking the primitive 
``$\send$ {\sc tag}$(m)$ $\sto~p_j$'', where {\sc tag} is the type
of the message and $m$ its content. To simplify the presentation, it is
assumed that a process can send messages to itself. A process $p_i$ receives 
a message by executing the primitive ``$\receive()$''.
The macro-operation $\broadcast$ {\sc tag}$(m)$ is  used as a shortcut for
``{\bf for each} $p_i \in \Pi$  {\bf do} $\send$ {\sc tag}$(m)$ $\sto~p_j$
{\bf end for}''. 
\vspace{0.5em}
  
\noindent
{\bf Failure model.}
Up to $t$ processes can exhibit a {\it Byzantine} behavior~\cite{PSL80}.
 A Byzantine process is a process that behaves
arbitrarily: it can crash, fail to send or receive messages, send
arbitrary messages, start in an arbitrary state, perform arbitrary state
transitions, etc. Moreover, Byzantine processes can collude 
to ``pollute'' the computation (e.g., by sending  messages with the same 
content, while they should send messages with distinct content if 
they were non-faulty). 
A process that exhibits a Byzantine behavior is called {\it faulty}.
Otherwise, it is {\it non-faulty}.  
Let us notice that, as each pair of processes is connected by a channel, 
no Byzantine process can impersonate another process.   
Byzantine processes can control the network by modifying
the order in which messages are received, but they cannot
postpone forever message receptions.  
\vspace{0.5em}

\noindent
{\bf Additional synchrony assumption.}
It it well-known that there is no consensus algorithm ensuring both
safety and liveness properties in fully asynchronous message-passing
systems in which even a single process may crash~\cite{FLP85}.  As the
crash failure model is less severe than the Byzantine failure model,
the consensus impossibility remains true if processes may commit
Byzantine failures.
To circumvent such an impossibility, and ensure the consensus termination
property, 
we enrich the model with additional synchrony
assumptions. 
%
It is assumed that after some finite time $\tau$,
there is an upper bound $\delta$ on message transfer and process computation delays.  This
eventual (or partial) synchrony assumption is denoted $\ES$.
\vspace{0.5em}

\noindent
{\bf Notations.}
The acronym ${\BAMP}[\emptyset]$ is used to denote the previous 
basic Byzantine Asynchronous Message-Passing computation model;
$\emptyset$ means that there is no additional assumption. 
The basic computation model strengthened with the additional constraint $t<n/3$
is denoted ${\BAMP}[t<n/3]$.
The latter computation model strengthened with the eventual synchrony constraint $\ES$
is denoted ${\BAMP}[t<n/3,\ES]$.

\section{Binary Byzantine Consensus}
\label{sec:byz-consensus}
\sloppy{In this section we propose a solution to the binary consensus using a weak coordinator that requires neither signatures, nor randomization.
%
For the sake of simplicity, we build the algorithm incrementally by first recalling the binary consensus problem, 
then presenting a safe binary consensus algorithm in the ${\BAMP}[t<n/3]$ model and finally presenting a safe and live consensus algorithm in the  ${\BAMP}[t<n/3,\ES]$ model.}




%

Let $\cal V$ be the set of values that can be proposed by a process to the consensus.
While  $\cal V$ can contain any number ($\geq 2$) of values
in multivalued consensus, it contains only two values in binary consensus, 
e.g., ${\cal V} =\{0,1\}$.
Assuming that
each non-faulty process proposes a value, the binary Byzantine consensus (BBC) problem is for 
each of them to
decide on a value in such a way that the following properties are
satisfied:
\begin{smallenum}
\item BBC-Termination. Every non-faulty process eventually decides on a value.
\item BBC-Agreement.   No two non-faulty processes decide on different values.
\item BBC-Validity.  If all non-faulty processes propose the same value, no
other value can be decided.
\end{smallenum}


\subsection{The Binary Value Broadcast Communication Abstraction}\label{ssec:bvbcast}

Our binary consensus algorithm relies on a binary value all-to-all 
communication abstraction, denoted BV-broadcast, originally introduced for randomized 
consensus~\cite{MMR15}, and restated in the appendix.
%

In a  BV-broadcast instance, each non-faulty process $p_i$  broadcasts 
a  binary value and obtains (BV-delivers) a set of binary values, stored in a local
read-only set variable denoted $\binvalues_i$.
This set, initialized to $\emptyset$, increases when new values are received.  
BV-broadcast is defined by the four following  properties:
\begin{itemize}[noitemsep]
\item BV-Obligation. 
If at least $(t+1)$ non-faulty  processes BV-broadcast the same value $v$, $v$  
is eventually added to the set $\binvalues_i$ of each non-faulty process $p_i$. 
\item BV-Justification. 
If $p_i$ is non-faulty and  $v\in \binvalues_i$,  $v$ has been 
BV-broadcast by a non-faulty process. 
\item BV-Uniformity. 
If a value $v$ is added to the set $\binvalues_i$ of a non-faulty process $p_i$, 
eventually  $v\in \binvalues_j$ at every non-faulty process $p_j$. 
\item BV-Termination. 
Eventually the set $\binvalues_i$ of  each non-faulty process $p_i$  is not
empty.  
\end{itemize}

The following property is an immediate consequence of the  previous properties. 
Eventually the sets  $bin\_values_i$ of  the  non-faulty processes $p_i$ 
(i)~become non-empty,  (ii)~become  equal,
(iii)~contain all the values broadcast by non-faulty processes, and
(iv)~never contain a value broadcast only by Byzantine processes.
However, no non-faulty process knows  when (ii) and (iii) occur.

%

\subsection{Local variables and message types}
Each process $p_i$ manages the following local variables.
\begin{smallenum}
\item $est_i$: local current estimate of the decided value.
It is initialized to the value proposed by $p_i$. 
\item  $r_i$:  local asynchronous round number, initialized to $0$.
\item $\binvalues_i[1..]$: 
array of binary values; $\binvalues_i[r]$ (initialized to $\emptyset$)
stores the   local output set filled by BV-broadcast associated with round $r$.
(This unbounded array can be replaced by a single local variable
$\binvalues_i$, reset to $\emptyset$ at the beginning of every round. 
We consider here an array to simplify the presentation.) 
\item $b_i$: auxiliary binary value.
\item $\values_i$: auxiliary set of values.
\end{smallenum}



The algorithm uses two message types, denoted {\sc est} and {\sc aux}.
Both are used in each round, hence they always appear with a round number.
\begin{smallenum}
\item
{\sc est}$[r]()$ is used at round $r$ by $p_i$ to BV-broadcast its
current decision estimate $est_i$. 
\item
{\sc  aux}$[r]()$ is used by $p_i$ to disseminate  its current value of
$\binvalues_i[r]$ (with the help of the $\broadcast()$ macro-operation).
\end{smallenum}

\subsection{A safe asynchronous binary Byzantine consensus algorithm}\label{sec:safe-bbc}
For the sake of simplicity, we first introduce a new leaderless algorithm ensuring BBC-Validity and
BBC-Agreement properties in the system model ${\BAMP}[t<n/3]$ but not BBC-termination.
%
%
The algorithm is depicted in Figure~\ref{algo-Bin-Byz-consensus-safe} and
provides the process $p_i$ with the operation
$\binpropose(v_i)$ to propose its initial value $v_i$.  Process $p_i$ proceeds in asynchronous rounds and decides value $v$ when invoking $\decide(v)$ at line~\ref{BYZ-safe-10}.

\begin{figure*}[ht!]
\centering{
\fbox{
\begin{minipage}[t]{150mm}
\footnotesize
\renewcommand{\baselinestretch}{2.5}
\resetline
\begin{tabbing}
aaaA\=aaA\=aaaA\=aaaaaaaaaA\kill

{\bf opera}\={\bf tion} ${\binpropose}(v_i)$ {\bf is}\\

\line{BYZ-safe-01} \> $est_i\leftarrow v_i$;
                     $r_i \leftarrow 0$;\\

\line{BYZ-safe-02} \> {\bf while} $(\ttrue)$  {\bf do}\\

\line{BYZ-safe-03} \>\> $r_i \leftarrow r_i+1$;\\

\line{BYZ-safe-04} \>\> ${\BVbroadcast}$  {\sc est}$[r_i](est_i)$; {\it \scriptsize  \hfill \color{gray}{// add to $\binvalues{[r_i]}$ upon $BV\_delivery$}} \\

\line{BYZ-safe-05} \>\> $\wait$ \big($\binvalues_i[r_i]\neq\emptyset$\big); \\

\line{BYZ-safe-06} \>\> ${\broadcast}$  {\sc aux}$[r_i](\binvalues_i[r_i])$;\\

\line{BYZ-safe-07} \>\>  $\wait$

\big(messages {\sc aux}$[r_i](b\_val_{p(1)})$, ..., 
               {\sc aux}$[r_i](b\_val_{p(n-t)})$ have been received \\

\>\>   $~~~~~~~~~~~~~~~~~~$  from $(n-t)$ different processes
               $p(x)$, $1\leq x\leq n-t$,  and their contents are\\

\>\>   $~~~~~~~~~~~~~~~~~~$
               such that $\exists$ a  non-empty set $\values_i$ where 
               (i)  $\values_i=\cup_{1\leq x\leq n-t} b\_val_{p(x)}$\\
\>\>   $~~~~~~~~~~~~~~~~~~$ and
               (ii)  $\values_i\subseteq \binvalues_i[r_i]$\big);\\    

\line{BYZ-safe-08} \>\> $b_i \leftarrow r_i \modulo 2$;\\

\line{BYZ-safe-09} \>\> {\bf if} \= $(\values_i=\{v\})$
     {\it \scriptsize  \hfill \color{gray}{// $\values_i$
                                       is a singleton whose element is $v$}}\\

\line{BYZ-safe-10} \>\>\>

{\bf then} \= $est_i\leftarrow v$; 

{\bf if} $(v=b_i)$   {\bf then} $\decide(v)$ if not yet done {\bf  end if};\\

\line{BYZ-safe-11} \>\>\> {\bf else} \> $est_i\leftarrow b_i$\\

\line{BYZ-safe-12} \>\> {\bf end if};\\

\line{BYZ-safe-13} \> {\bf end while}. \\~\\

\line{BYZ-safe-14} \>{\bf when}  {\sc b-val}$[r](v)$ is {\bf BV-delivered} by ${\BVbroadcast}[r]$ {\bf do} \\
  \>\> $\binvalues_i[r] \leftarrow \binvalues_i[r] \cup \{v\}$;

\end{tabbing}
\normalsize
\end{minipage}
}
\caption{A safe algorithm for the binary Byzantine consensus in ${\BAMP}[t<n/3]$}
\label{algo-Bin-Byz-consensus-safe} 
}
\vspace{-1em}
\end{figure*}

After it has deposited its binary proposal in $est_i$ (line~\ref{BYZ-safe-01}),
each non-faulty process $p_i$ enters a sequence of asynchronous rounds.
During a round $r$, each non-faulty process $p_i$ proceeds
in three phases.\\

\noindent
{\bf Phase 1: Binary value broadcast to filter out the values of Byzantine processes.}
Process $p_i$ first
  progresses to the next round, and binary value broadcasts (BV-broadcast) its current estimate
  (line~\ref{BYZ-safe-04}). 
%

  At each process $p_i$, within the ${\BVbroadcast}()$ algorithm, after receiving the same value from $t+1$ processes, process $p_i$ then rebroadcasts this value.
  Each process $p_i$ BV-delivers a value $v$ by adding it to its $\binvalues_i$ set only if it receives $v$ from $2t+1$ distinct processes.  Eventually the sets $bin\_values$ of all non-faulty processes 
become non-empty, equal, and contain exclusively all values broadcast by non-faulty processes~\cite{CGLR18}.  
When a value is BV-delivered it is then added to $\binvalues_i[r]$ (line \ref{BYZ-safe-14}).
%
  Then $p_i$ waits until its set
  $\binvalues_i[r]$ is not empty (let us recall that, when
  $\binvalues_i[r]$ becomes non-empty, it has not necessarily its
  final value). \\

\noindent
{\bf Phase 2: Exchanging estimates to converge to an agreement.}
This second phase runs between line~\ref{BYZ-safe-06} and line~\ref{BYZ-safe-07}).
In this phase, $p_i$ broadcasts normally 
  a message {\sc aux}$[r]()$ whose
  content is $\binvalues_i[r]$   (line~\ref{BYZ-safe-06}).
  Then, $p_i$ waits until it has received a set of
  values $\values_i$  satisfying the two following properties.
  \begin{smallenum}
\item The values in $\values_i$ come from the messages {\sc aux}$[r]()$
  of at least $(n-t)$ different processes.
\item $\values_i\subseteq \binvalues_i[r]$. 
Thanks to the BV-broadcast that filters out Byzantine value,
  even if Byzantine
  processes send fake messages {\sc aux}$[r]()$ containing values
  proposed only by Byzantine processes,
  $\values_i$ will contain only
  values broadcast by non-faulty processes.
\end{smallenum}
Hence, at any round $r$, after line~\ref{BYZ-safe-07}, 
$\values_i\subseteq \{0,1\}$ and contains only values
BV-broadcast at line~\ref{BYZ-safe-04} by non-faulty processes.\\

\noindent
{\bf Phase 3: Deciding upon estimate convergence to round number modulo 2.}
The third phase runs between line~\ref{BYZ-safe-08} and line~\ref{BYZ-safe-12}.
  This phase is a purely local computation phase, during which (if not yet done)
  $p_i$ tries to decide the value $b=r\modulo 2$ (lines~\ref{BYZ-safe-08}
  and~\ref{BYZ-safe-10}), depending on the content of $\values_i$.
\begin{smallenum}
  \item
    If $\values_i$ contains a single element $v$ (line~\ref{BYZ-safe-09}),
    then $v$ becomes $p_i$'s new estimate. Moreover, $v$ is a
    candidate for the consensus decision. To ensure BBC-Agreement, $v$ can be decided
    only if $v=b$. The decision is realized by the statement $\decide(v)$
    (line~\ref{BYZ-safe-10}). 
\item
  If  $\values_i=\{0,1\}$, then $p_i$ cannot decide. As both values have been
  proposed by non-faulty processes, to entail convergence to agreement,
  $p_i$ selects one of them ($b$, which is the same at all non-faulty
  processes for the same round) as its new estimate  (line~\ref{BYZ-safe-11}). 
 \end{smallenum} 
Let us observe that the invocation of $\decide(v)$ by $p_i$
does not terminate the participation of $p_i$ in the algorithm,
namely $p_i$ continues looping forever. This is because a deciding process
may need to help other processes converging to the decision in the two subsequent rounds.  
This algorithm can be modified to avoid this infinite loop,
but to preserve the simplicity in the presentation, we postpone a deterministic
terminating solution to Section~\ref{ssec:live-bbc}.
The proof of correctness of algorithm~\ref{algo-Bin-Byz-consensus-safe} is deferred to the appendix.

\subsection{Psync: Safe and Live Consensus in ${\BAMP}[t<n/3,\ES{}]$}
\label{ssec:live-bbc}

We now present \emph{Psync}, an algorithm solving the binary Byzantine consensus problem in the ${\BAMP}[t<n/3,\ES{}]$ model. 
Similar to the safe algorithm (Section~\ref{sec:safe-bbc}), Psync does not use signatures or randomization and has the following additional characteristics: 
\begin{itemize}
\item Psync is time optimal~\cite{FL82} in that it terminates in $O(t)$ message delays.
\item When all non-faulty processes propose the same value, 
Psync terminates in O(1) message delays, even under asynchrony. 
\item Psync does not wait for a message from its coordinator and does not 
need recovery.
\end{itemize}

%


The Psync algorithm is presented in Figure~\ref{algo-Bin-Byz-consensus-safe-live} as an extension
of the safe algorithm in Figure~\ref{algo-Bin-Byz-consensus-safe}, with new and modified lines prefixed with ``New'' and 
``M-'', respectively. Lines prefixed by ``Opt'' are optional optimizations.
%
In addition to the use of local timers, to eventually
benefit from the $\ES$ assumption, the algorithm uses a {\it weak coordinator}:
the weak coordinator of
round $r$ is the process $p_i$ such that $i=((r-1) \mod n)
+1$.
Note that this new round coordinator is only 
used to help agreement by suggesting a value and thus differs from the classic coordinator~\cite{CT96,DLS88}.


\begin{figure*}[ht!]
\centering{
\fbox{
\begin{minipage}[t]{150mm}
\footnotesize
\renewcommand{\baselinestretch}{2.5}
\resetline
\begin{tabbing}
aaaA\=aaA\=aaaA\=aaaA\=aaaaaaaaaA\kill

{\bf opera}\={\bf tion} ${\binpropose}(v_i)$ {\bf is}\\

(\ref{BYZ-safe-01}) \> $est_i\leftarrow v_i$;
                     $r_i \leftarrow 0$;\\
                     \> $timeout_i \leftarrow 0$;\\

(\ref{BYZ-safe-02}) \> {\bf while} $(\ttrue)$  {\bf do}\\

(\ref{BYZ-safe-03}) \>\> $r_i \leftarrow r_i+1$;\\

(Opt1) \>\> \color{blue}{{\bf if} $(est_i = -1)$ {\bf then} $est_i \leftarrow 1$;} \color{gray}{{\it \scriptsize  \hfill {// ``fast-path''  for round 1, only used in the reduction in Sect.~\ref{sec:multi-to-bin}}}} \\
(\ref{BYZ-safe-04})        \>\>~~ {\bf else} ${\BVbroadcast}$  {\sc est}$[r_i](est_i)$;\\
        \>\> {\bf end if};\\
(New1)  
        \>\> $\wait$ \big($\binvalues_i[r_i]\neq\emptyset$\big);\\
        \>\> $timeout_i \leftarrow  timeout_i + 1$; 
           $\set$ $timer_i$  $\sto$  $timeout_i$;\\

(New2)  \>\> $\coord_i \leftarrow  ((r_i-1) \mod n) +1$;\\
        \>\> {\bf if} \= $(i=\coord_i)$ 
{\bf then}\\
            
     \>\>\> $\{w\} = \binvalues_i[r_i]$;
\color{gray}{{\it \scriptsize  \hfill {// $w$ is the first value to enter $\binvalues_i[r_i]$}}}\\
     \>\>\> ${\broadcast}$  \coordval$[r_i](w)$\\
     \>\> {\bf end if};\\
                 
(M-\ref{BYZ-safe-05}) \>\> $\wait$
 \big($(\binvalues_i[r_i]\neq\emptyset)\wedge (timer_i \mbox{ expired})$\big);\\

(New3)  \>\> {\bf if} \= \big(({\coordval}$[r_i](w)$ received from $p_{\coord_i}$)
                               $\wedge$ $(w\in \binvalues_i[r_i])$\big)\\  
     \>\>\> {\bf then} \=$aux_i\leftarrow \{w\}$\\
     \>\>\> {\bf else} \>$aux_i\leftarrow \binvalues_i[r_i]$\\
     \>\> {\bf end if};\\
 
(M-\ref{BYZ-safe-06}) \>\> ${\broadcast}$  {\sc aux}$[r_i](aux_i)$;\\

     (New4)  \>\> $\wait$ \big(a message {\sc aux}$[r_i]()$ has been received from $(n-t)$ different processes\big); \\
     \>\>$\set$ $timer_i$  $\sto$  $timeout_i$;\\
     
(M-\ref{BYZ-safe-07}) \>\>  $\wait$

\big((messages {\sc aux}$[r_i](b\_val_{p(1)})$, ..., 
               {\sc aux}$[r_i](b\_val_{p(n-t)})$ have been received \\

\>\>   $~~~~~~~~~~~~~~~~~~$  from $(n-t)$ different processes
               $p(x)$, $1\leq x\leq n-t$,  and their contents are\\

\>\>   $~~~~~~~~~~~~~~~~~~$
               such that $\exists$ a  non-empty set $\values_i$ where 
                (i)  $\values_i=\cup_{1\leq x\leq n-t} b\_val_{p(x)}$\\
\>\>   $~~~~~~~~~~~~~~~~~~$ and
              (ii)  $\values_i\subseteq \binvalues_i[r_i]$)
              $\wedge$ ($timer_i$ expired)\big);\\

(New5)  \>\> {\bf if} \=
(when considering the whole set of the  messages {\sc aux}$[r_i]()$ received,
several sets \\
\>\>\> $~values1_i$, $values2_i$, ... satisfy the previous wait
   predicate) $\wedge$ (one of them is $aux_i$) \\

\>\>  $~~~~~~~~~~~~~~~~~~$
            {\bf then}  $\values_i \leftarrow aux_i$ {\bf end if};    \color{gray}{{\it \scriptsize  \hfill {//
                           $values_i$ is either defined here or at line M07}}} \\ 

(\ref{BYZ-safe-08}) \>\> $b_i \leftarrow r_i \modulo 2$;\\

(\ref{BYZ-safe-09}) \>\> {\bf if} \= $(\values_i=\{v\})$
     \color{gray}{{\it \scriptsize  \hfill {// $values_i$
                                    is a singleton whose element is $v$}}}\\
     
(\ref{BYZ-safe-10}) \>\>\>

{\bf then} \= $est_i\leftarrow v$; 

{\bf if} $(v=b_i)$   {\bf then} $\decide(v)$ if not yet done {\bf  end if};\\

(\ref{BYZ-safe-11}) \>\>\> {\bf else} \> $est_i\leftarrow b_i$\\

(\ref{BYZ-safe-12}) \>\> {\bf end if};\\

(Opt2) \>\> \color{blue}{{\bf if} $($decided in round $r_i)$ {\bf then}}
       \color{gray}{{\it \scriptsize  \hfill {//
                           the following are termination conditions}}}\\
\>\>\> \color{blue}{{\bf wait until} $(\binvalues_i[r_i] = \{0,1\})$ 
       \color{gray}{{\it \scriptsize  \hfill {//
                           only go to the next round when necessary}}}}\\ 
\>\> \color{blue}{{\bf else if} $($decided in round $r_i-2)$ {\bf then ${\sf halt}$ end if;} 
       \color{gray}{{\it \scriptsize  \hfill {//
                           everyone has decided by now}}}}\\ 
 \>\>  \color{blue}{{\bf end if};}\\

(\ref{BYZ-safe-13}) \> {\bf end while}.
       
\end{tabbing}
\normalsize
\end{minipage}
}
\caption{A safe and live  algorithm for the binary Byzantine consensus in
  ${\BAMP}[t<n/3,\ES]$; line (Opt1) is an optimization only applied in the multivalued reduction
  presented in Section \ref{sec:multi-to-bin}; line (Opt2) is a mechanism
  to prevent unnecessary rounds from being executed}
\label{algo-Bin-Byz-consensus-safe-live} 
}\vspace{-1em}
\end{figure*}

\noindent
{\bf Additional local variables and message type.}
In addition to $est_i$, $r_i$, $\binvalues_i[r]$, and $\values_i$,
each process $p_i$ manages the following local variables.
\begin{smallenum}
\item $timer_i$ is a local timer, and $timeout_i$ a timeout value, both
  used to exploit the assumption $\ES$. 
\item $coord_i$ is the index of the current weak round coordinator.
\item $aux_i$ is an auxiliary set of values, used to store the value
  (if any) that the current weak coordinator strives to impose as decision
  value.
\end{smallenum}
The weak coordinator of round $r$, uses the message type {\coordval}$[r]()$ to
broadcast the value it suggests for decision.   \\

\noindent
{\bf Description of the extended algorithm.}
We now list the new and modified lines that were added in
Figure~\ref{algo-Bin-Byz-consensus-safe-live}.
\begin{smallenum}
\item At line New1, $p_i$ waits until a value enters $\binvalues$, then
  sets its local timer, whose expiry is used in the predicate of
  line~M-\ref{BYZ-safe-05}.  The timeout value is initialized before
  entering the loop, and then increased at every round.
\item Line Opt1 is an optimization only used along
  with the reduction to multivalued consensus presented in
  Section \ref{sec:multi-to-bin}.
\item Line New4 waits until $(n-t)$ {\sc aux}$[r]()$ messages are received from
  different processes before reseting
  the timer, whose expiry is
  used in the predicate of the modified line~M-\ref{BYZ-safe-07}. 
\item Lines New2, New3, M-\ref{BYZ-safe-06}, and New5 realize a
  mechanism that allows the current weak coordinator (whose value is computed
  on line New2) to try to impose
  the first value that enters into its
  $\binvalues$ set as the decided value.
   Combined
  with the fact that there is a time after which the messages
  exchanged by the non-faulty processes are timely, this ensures that
  there will be a round during which the non-faulty processes will have a
  single value in their sets $\values_i$, which entails their decision.
\item Modified lines M-\ref{BYZ-safe-05} and M-\ref{BYZ-safe-07}: 
  addition of the timer expiration in the predicate considered at the
  corresponding line.
\item Line Opt2 is an optional optimization to minimize the amount of extra
  rounds processes need to execute after deciding.
  The first condition ({\bf wait until} $(\binvalues_i[r_i] = \{0,1\})$)
  ensures that, after decision, a process only continues to the next round
  if some other non-faulty process did not decide in the current round.
  As this can only happen if both $0$ and $1$ enter $\binvalues$,
  the process will not move on to the next round until this is true.
  The second condition, ({\bf if} $($decided in round $r_i-2)$),
  halts the process $2$ rounds after it has decided, as all non-faulty processes
  are guaranteed to have decided by this round.
\end{smallenum}

The aforementioned modifications exploit the weak coordinator that 
only helps resolving disagreement by broadcasting 
a value that all non-faulty adopt, as opposed to leaders or
classic (strong) coordinators~\cite{CT96,DLS88}.
%
%
To this end:
\begin{smallenum}
\item The weak coordinator $p_k$ broadcasts the message
  {\coordval}$[r_i](w)$, where $w$ is the first value that enters its
  $\binvalues$ set (line New2). If $p_k$ is non-faulty, the timeout
  values of the non-faulty processes are big enough, and there is a bound
  on message transfer delays, so that all non-faulty processes will receive it
  before their timer expiration at line  M-\ref{BYZ-safe-05}.
\item Then, assuming the previous item, all non-faulty processes set
  $aux_i$ to $\{w\}$ (line New3), and broadcast it (line
  M-\ref{BYZ-safe-06}).  The predicate $w\in \binvalues_i[r_i]$ is
  used to prevent a Byzantine coordinator to send fake values that
  would foil non-faulty processes.
\item Finally, all the
  non-faulty processes  will receive the message {\sc aux}$[r_i](\{w\})$ from
  $(n-t)$ different processes,
  and, by line New5, will set $\values_i=\{w\}$. 
  This entails their decision during the round $(r+1)$ or $(r+2)$.
\end{smallenum}
%
%
To ensure that slow processes catch up to faster processes that have reached later rounds, once a process
has received at least $t+1$ messages belonging to a round $r$, the process does wait for timeouts in rounds less than $r$.
In the presence of $\ES$, this ensures that all non-faulty processes eventually execute synchronous rounds.
The proof of liveness of algorithm~\ref{algo-Bin-Byz-consensus-safe-live} is deferred to the appendix.
\section{DBFT: From Binary Byzantine Consensus to Blockchain Consensus}
\label{sec:multi-to-bin}

This section presents a \emph{Democratic Binary Fault Tolerant} algorithm, called DBFT.
It relies on a reduction from the binary Byzantine consensus Psync to the multivalue consensus and is also time optimal, resilience optimal and does not use classic (strong) coordinator, which means that it 
does not wait for a particular message. In addition, it finishes in only 4 messages delays in the good case, when all non-faulty processes propose the same value.
%

We consider a variant of the classical Byzantine
consensus problem, called the \emph{Validity Predicate-based Byzantine
Consensus} (denoted VPBC). 
Its
validity requirement 
relies on an
application-specific $\valid()$ predicate that is used by blockchains to indicate whether a value
is \emph{valid}.
Assuming that
each non-faulty process proposes a valid value, each of them has to
decide on a value in such a way that the following properties are
satisfied.
\begin{smallenum}
\item VPBC-Termination. Every non-faulty process eventually decides on a value.
\item VPBC-Agreement.   No two non-faulty processes decide on different values.
\item VPBC-Validity.  A decided value is valid, i.e., it satisfies the predefined predicate
denoted $\valid()$, and if all non-faulty processes propose the same value $v$ then they decide $v$.
\end{smallenum}


This definition 
generalizes the classical definition of
Byzantine consensus, which does not include the predicate $\valid()$.
This predicate is introduced to take into account the distinctive
characteristics of consortium blockchains,
and possibly other specific Byzantine consensus problems.
In the context of consortium blockchains, a proposal is not valid if
it does not contain an appropriate hash of the last block added to the
Blockchain or contains invalid transactions.
There exist similar problem definitions whose validity also
relies on the notion of a predicate. The validated Byzantine
consensus~\cite{CGR11} differs in that the same valid value proposed
by non-faulty processes has to be decided if all processes are non-faulty.
The asynchronous
Byzantine agreement~\cite{Kur00} defines a legal value similar to our
valid value, however, its validity does not require a legal value to
be decided if multiple ones exist, while we require that
any decided value must be valid. 
A probabilistic variant~\cite{CKPS01} required that the decided value
be one of the proposed values, something we do not require. 


\begin{figure}[htbp]
\centering{
\fbox{
\begin{minipage}[t]{150mm}
\footnotesize
\renewcommand{\baselinestretch}{2.5}
\resetline
\begin{tabbing}
aA\=aaaaaaaaaaaA\=aaaaaaaaaaaaA\kill

{\bf opera}\={\bf tion} $\mvpropose(v_i)$ {\bf is}\\

\line{MV-BYZ-01} \>   ${\sf  RB\_broadcast}$  {\sc val}$(v_i)$;\\

\line{MV-BYZ-02} \>
 {\bf repeat} \= {\bf if} \big($\exists~k:
                        (\proposals_i[k]\neq \bot) \wedge$ \\
 \>\> $(\BINCONS[k].\binpropose()$ not invoked)\big)\\

\line{MV-BYZ-03} \> \>  $~~~~$
          {\bf then} invoke $\BINCONS[k].\binpropose(-1)$ {\bf end if};\\

\line{MV-BYZ-04}  \>
 {\bf until}  $(\exists \ell:~ \bindec_i[\ell] = 1)$ {\bf end repeat};\\

\line{MV-BYZ-05} \> {\bf for each} $k$ 
s.t. $\BINCONS[k].\binpropose()$ not yet invoked\\

\line{MV-BYZ-06} \> \>
{\bf do} invoke $\BINCONS[k].\binpropose(0)$ {\bf end for};\\
 
\line{MV-BYZ-07} \> $\wait$ $(\bigwedge_{1\leq x\leq n} \bindec_i[x]\neq \bot)$;\\

\line{MV-BYZ-08} \>
        $j \leftarrow \mmin \{x \mbox{ such that } \bindec_i[x]=1\}$;\\

\line{MV-BYZ-09}  \> $\wait$ $(\proposals_i[j]\neq\bot)$;\\

\line{MV-BYZ-10}  \> $\decide(\proposals_i[j])$.\\~\\

\line{MV-BYZ-11}
{\bf when}  {\sc val}$(v)$ {\bf is RB-delivered from} $p_j$ {\bf do} \\
\> {\bf if} $\valid(v)$ {\bf then} \\
         \>\>$proposals_i[j] \leftarrow v$; \\
         \>\> {\bf BV-deliver} {\sc b-val}$[1](1)$ to $\BINCONS[j]$ {\bf end if}.\\~\\

\line{MV-BYZ-12} 
     {\bf when} $\BINCONS[k].\binpropose()$ {\bf decides a value} $b$ \\
       \> {\bf do} $\bindec_i[k] \leftarrow b$. 
         
\end{tabbing}
\normalsize
\end{minipage}
}
\caption{From multivalued to binary Byzantine consensus
                         in $\BAMP[t<n/3,\mbox{BBC}]$}
\label{algo:reduction-multi-to-bin} 
}
\vspace{-1em}
\end{figure}



\noindent
{\bf Binary consensus objects.}
The
processes cooperate with an array of binary Byzantine consensus
objects denoted $\BINCONS[1..n]$.  The instance $\BINCONS[k]$ allows
the non-faulty processes to find an agreement on the value proposed by
$p_k$. This object is implemented with the binary Byzantine consensus
algorithm presented in Section~\ref{ssec:live-bbc}.
%
To simplify the presentation, we consider that a process
$p_i$ launches its participation in $\BINCONS[k]$ 
by invoking  $\BINCONS[k].\binpropose(v)$, where $v\in\{0,1\}$. 
Then, it executes the corresponding code in a specific thread, 
which eventually returns the value decided by $\BINCONS[k]$.\\

\noindent
{\bf Local variables.}
Each process $p_i$ manages the following local variables; $\bot$
denotes a default value that cannot be proposed by a
(faulty or non-faulty) process.

\begin{smallenum}
\item
An array $\proposals_i[1..n]$ initialized to $[\bot, \cdots,\bot]$.
The aim of $\proposals_i[j]$ is to contain the value proposed by $p_j$. 
\item
An array $\bindec_i[1..n]$ initialized to $[\bot, \cdots,\bot]$.
The aim of $\bindec_i[k]$ is to contain the value ($0$ or $1$)  decided by
the binary consensus object $\BINCONS[k]$. 
\end{smallenum}

\noindent
{\bf The algorithm.}
The algorithm reducing from the binary Byzantine consensus to multivalue
Byzantine consensus is described in Figure~\ref{algo:reduction-multi-to-bin} and is 
 similar to an existing reduction~\cite{BKR94}, except that
it 
combines the reliable broadcast, RB-broadcast~\cite{B87}, restated in the appendix,
with our binary consensus messages to finish
in 4 message delays in the good case.
Initially, a process invokes the operation $\mvpropose(v)$, where
$v$ is the value it proposes to the multivalued consensus.  
Process $p_i$ executes four phases. 
\vspace{0.1cm}\\
\noindent
{\bf Phase 1:}
$p_i$ disseminates its value (lines~\ref{MV-BYZ-01} and~\ref{MV-BYZ-11}).
Process $p_i$ first sends its value to all the processes by invoking
the RB-broadcast operation (line~\ref{MV-BYZ-01}). 
If a process RB-delivers a valid value $v$ RB-broadcast by a process $p_j$, 
then the process stores it in $proposals_i[j]$ and BV-delivers
$1$ directly to round one of instance $\BINCONS[j]$ (line~\ref{MV-BYZ-11}),
placing $1$ in its $\binvalues_i$ for that instance.
\vspace{0.1cm}\\
\noindent
{\bf Phase 2:}
Process $p_i$ starts participating in a first set of binary
consensus instances (lines~\ref{MV-BYZ-02}-\ref{MV-BYZ-04}).
It enters a loop in which it starts participating in the binary
consensus instances.
Process $p_i$ invokes a binary consensus 
instance $k$ with value $-1$ for each value RB-broadcast by process $p_k$ that $p_i$ RB-delivered.
$-1$ is a special value that allows the binary consensus to skip the
$\BVbroadcast$ step (line~(Opt1)) and immediately send an {\sc aux} message with value $1$,
allowing the binary consensus to terminate with value $1$ in a single message delay.
(Note that the timeout of the first round is set to $0$ so the binary consensus proceeds
as fast as possible.)
The direct delivery of $1$ into $\binvalues$ is possible due to an overlap in the properties of $\BVbroadcast$ and
RB-broadcast, allowing us to skip a message step of our binary consensus algorithm.
In other words, 
all non-faulty
processes will RB-deliver the proposed value, and as a result will also BV-deliver $1$.
This loop stops
as soon as $p_i$ discovers a  binary consensus instance
$\BINCONS[\ell]$ in which $1$ was decided (line~\ref{MV-BYZ-04}).
(As all non-faulty processes will only have $1$ in their $\binvalues$ until an instance terminates,
the first instance to decide $1$ will terminate in one message delay following
the RB-delivery.)
\vspace{0.1cm}\\
\noindent
{\bf Phase 3:} $p_i$ starts participating in all other binary consensus instances
(lines~\ref{MV-BYZ-05}-\ref{MV-BYZ-06}).
After it knows a binary consensus  instance decided 1, 
$p_i$ invokes with $\binpropose(0)$ all the binary consensus
instances  $\BINCONS[k]$ in which it has not yet participated.
Let us notice that it is possible that, for some of these
instances $\BINCONS[k]$, no process has RB-delivered a value from 
the associated process $p_k$. The aim of these consensus participation
is to ensure that all  binary consensus instances eventually terminate. 
\vspace{0.1cm}\\
\noindent
{\bf Phase 4:}
$p_i$ decides a value (lines~\ref{MV-BYZ-07}-\ref{MV-BYZ-10}
and~\ref{MV-BYZ-12}).
Process $p_i$ considers the first (according to the process index order) 
among the successful binary consensus objects, i.e., the ones that
returned $1$ (line~\ref{MV-BYZ-08}).
Let $\BINCONS[j]$ be this
binary consensus object. As the associated decided  value is $1$,
at least  one non-faulty process proposed $1$, which means that it RB-delivered
a value from the process $p_j$ (lines~\ref{MV-BYZ-02}-\ref{MV-BYZ-03}).
Observe that
this value is
eventually RB-delivered by every non-faulty process. Consequently,
$p_i$ decides it (lines~\ref{MV-BYZ-09}-\ref{MV-BYZ-10}).
Notice that as soon as the binary consensus instance with the
smallest process index terminates
with $1$, the reduction can return as soon as the associated
value is RB-delivered.
This is due to the observation that the values associated with the larger indices
will not be used.  \\
\vspace{-0.1cm}

\noindent
{\bf Complexity.}
This eager termination allows the consensus algorithm to terminate in $4$
message delays in the good scenario, i.e.,
$3$ message delays to execute the reliable broadcast and $1$ to complete
the binary consensus by skipping the $\BVbroadcast$ step.
In this case the reliable broadcast and binary consensus each have $O(n^2)$ message complexity
for a total of $O(n^3)$ including all $n$ instances.
In the case of faulty processes or asynchrony the algorithm will need
at least $3$ additional message delays for binary consensus
instances to terminate with $0$.


\begin{theorem}
\label{theorem-multivalued-consensus}
The algorithm described in Figure~{\em\ref{algo:reduction-multi-to-bin}}
implements the multivalued Byzantine consensus ({\em VPBC}) in the system model
${\BAMP}[t<n/3,\mbox{\em BBC}]$.
\end{theorem}
The proof of correctness of DBFT is deferred to the appendix.

\section{Conclusion}
\label{sec:conclusion}
To conclude, our weak coordinator based Byzantine consensus is time optimal, resilience optimal, does not rely on randomization or signatures
and improves over the randomized Byzantine consensus algorithms~\cite{MMR14,MMR15}
by terminating faster in various geo-distributed experiments.
We presented how it can be used for consortium blockchains by generalizing the Byzantine consensus problem and presenting a solution that combines an existing reduction with our 
binary Byzantine consensus algorithm. 

DBFT is now at the heart of the Red Belly Blockchain, a fast permissioned blockchain. Future work involves extending this permissioned blockchain into a public blockchain using DBFT for reconfiguration to periodically change at runtime the subset of machines running the consensus, similar to Solida~\cite{AMN17} but without proof-of-work.\\

\noindent
{\bf Acknowledgments.}
We wish to thank Christian Cachin and Seth Gilbert for their constructive feedback on earlier versions of this paper and for pointing us to relevant papers.

\newpage

\appendix

\section{Experiments on 100 VMs on Distinct Continents}
\label{sec:expe}

In this section, we evaluate the performance of our consensus algorithm against a randomized consensus applied to blockchains on 100 Amazon machines located  in 5 distinct data centers across different continents.

\subsection{Experimental setup}
To measure the performance of our consensus algorithm in a real network setting, we deployed our binary consensus algorithm
called ``Psync'' on 100 machines distributed across different continents. 

To implement point-to-point reliable channels over the Internet, we implemented secure channels using TLS on top of TCP/IP. Note that TLS uses a public key cryptosystem (and signatures) only to exchange secret keys, but no signatures are used by our consensus algorithm. Note that the
 Red Belly Blockchain builds upon the same combination of DBFT and TLS by storing the necessary certificates in its blocks~\cite{Gra17}.

For the sake of comparison, we also implemented the randomized binary Byzantine consensus algorithm from Most\'{e}faoui et al.~\cite{MMR14}, called ``Coin'', as a baseline. Coin terminates in $O(1)$ rounds in expectation and is at the heart of the HoneyBadger permissioned blockchain~\cite{MXC16} but requires a fair scheduler~\cite{MMR15}.
Our implementation reuses the common coin implementation of HoneyBadger~\cite{MXC16}  that consists of a one step message exchange and threshold signatures.
%
All 100 machines are c4.xlarge of Amazon EC2 equipped with an Intel Xeon E5-2666 v3  with 4 vCPUs,
7.5\,GiB RAM, and ``moderate'' network performance.

We set the timeouts of Psync to be null in the first $t$ rounds before incrementing exponentially. 
We implemented 
reliability using sequence numbers and negative acknowledgments at the application level.
All consensus decisions are stored to disk in an append only log.
Results are taken as the average of 100 instances of consensus.


\begin{figure}
\includegraphics[scale=0.55]{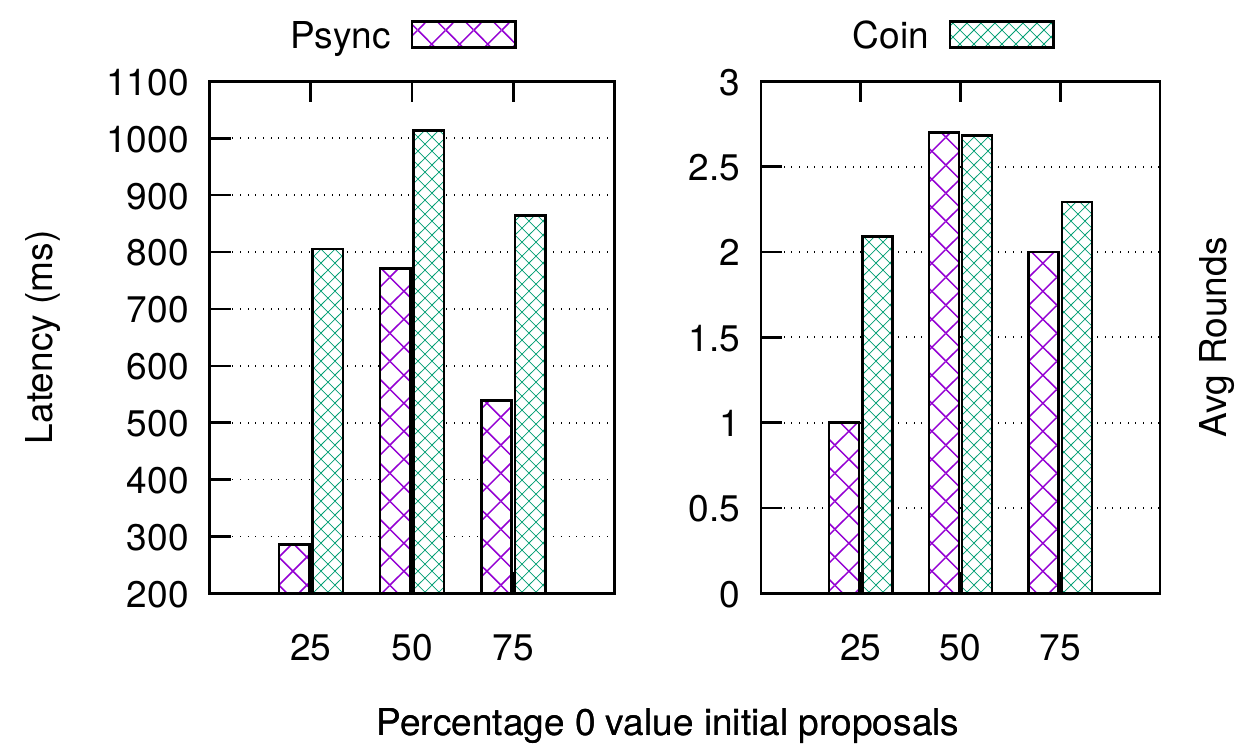}\hspace{0.5em}
\includegraphics[scale=0.55]{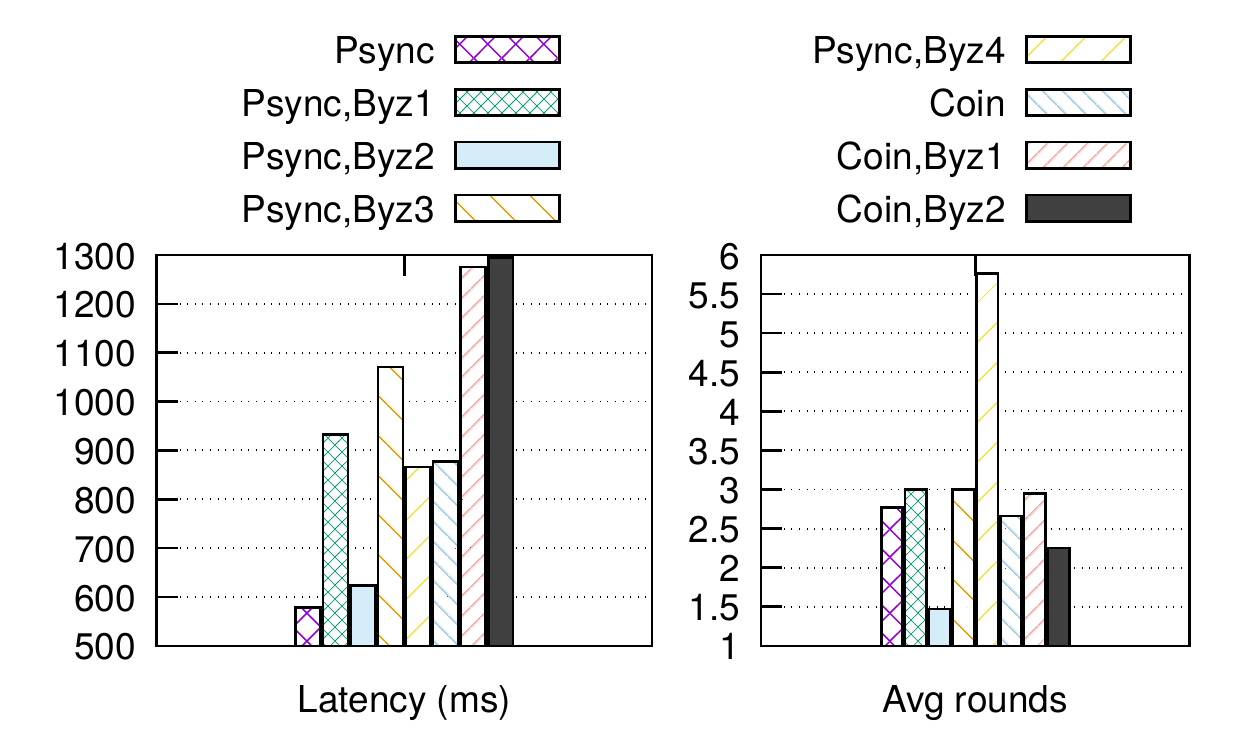}
\caption{Geo-distributed experiments of latency and average number of rounds for our deterministic binary Byzantine consensus ``Psync'' and the randomized binary Byzantine consensus ``Coin'' with 4 different Byzantine attacks (Byz1, ..., Byz4): {\bf (left)} with varying levels of initial disagreement, {\bf (right)} with random initial proposals.\label{fig:comparison}}
\vspace{-1em}
\end{figure}


\subsection{Geo-distributed experiments between 5 datacenters}
Figure~\ref{fig:comparison} compares the average latency and number of rounds needed to terminate Psync and Coin
in 5 Amazon datacenters, 3 in the US (Oregon, Northern California, and Ohio) and
2 in Europe (Ireland and Frankfurt). Our ping latency across continents is between 91\,ms and 164\,ms and within one continent between 22\,ms and 71\,ms.
In Figure \ref{fig:comparison}(left) 
the x-axis denotes the approximate percentage of processes that have an initial proposal of $0$ (others proposing $1$). 
Psync terminates in at most three rounds on average.

Given that Psync is designed to terminate with $1$ in the first round and $0$ in the second round,
the best performance is reached when the majority of proposals are $1$.
In all cases the latency of Psync is lower than Coin due to the coin needing an extra message step, additional computation complexity,
and randomness.

\subsection{Tolerance to various Byzantine attacks}
Figure \ref{fig:comparison}(right) compares the algorithms with
the following Byzantine behaviors: (Byz1) Byzantine processes flip the binary values of their messages; (Byz2) Byzantine processes are mute;
(Byz3) extends Byz1 with Byzantine coordinators that send random binary values in their {\coordval} messages; (Byz4) Byzantine processes form a coalition to limit the progress within rounds by
sending their own messages without waiting so they can be processed before others.
Both Byz3 and Byz4 are specific to Psync.

More precisely, Byz4 mimics a behavior where the coordinator is faulty to limit progress during rounds by trying
to have \textit{(i)}~no non-faulty processes to decide in round $r$ and \textit{(ii)}~have two non-faulty processes starting round $r+1$ with distinct estimates.  To this end, the faulty nodes start the round by broadcasting both $1$ and $0$ in their BV-broadcast. Then, the Byzantine coordinator sends a message $\coordval$
with $\neg(r \modulo 2)$ to all non-faulty nodes. Finally, Byzantine nodes instantly send
{\sc  aux} message with value $\neg(r \modulo 2)$
to a single node and send
{\sc  aux} message with value $(r \modulo 2)$
to the remaining nodes.
Faulty nodes in Byz4 have the power to send their messages instantly and to observe the messages received at non-faulty nodes, giving
them more power to delay termination.
They do not control the speed or order of messages from non-faulty nodes.

In Psync, the Byzantine processes are chosen as the first $t$ coordinators.
Coin has the highest latency with the Byzantine behaviors, but its number of rounds is least affected.
Byzantine behavior Byz3 is the slowest to terminate for Psync because it allows Byzantine processes to force the most disagreement.
While theoretically Byz4 could always prevent termination in the first $t$ rounds,
the average number of rounds is only increased to $6$ (but has a maximum of $35$).
This is due to the fact that they do not control the speed of messages of non-faulty processes in the network
preventing the non-terminating case.
Furthermore, given that the Byzantine processes have to act fast to ensure their messages are processed first,
the average latency is lower than Byz1 and Byz2.

\subsection{Detailed description of Byzantine behavior Byz4}
\label{sec:expe-byz4}
In the presence of a faulty coordinator it is possible to execute repeated rounds in which there
is no termination, behavior Byz4 tries to capture this behavior.
Note that we allow Byzantine messages to be delivered instantly by computing
them directly at the non-faulty nodes when needed.
We will now describe the Byz4 behavior.
Assume we are in a round $r$.
There are two main things we need to ensure:
(i)~no non-faulty process decides in round $r$
(ii)~at least one non-faulty node must start round $r+1$ with an estimate
of $0$ and another start with the estimate of $1$.

To ensure (i) we need (a) $\neg (r \modulo 2)$ to enter $\binvalues$
of non-faulty nodes and (b) no node must receive $n-t$ {\sc  aux}
messages with value $(r \modulo 2)$.
Then to ensure (ii) we need (c) both $0$ and $1$ to enter $\binvalues$
of non-faulty nodes, (d) at least one node must receive
receive $n-t$ {\sc  aux} messages with value $\neg(r \modulo 2)$,
and (e) at least one node must receive
receive an {\sc  aux} messages with value $(r \modulo 2)$.

Thus, Byzantine nodes start the round
by broadcasting both $1$ and $0$ in their BV-broadcast
to ensure (a) and (c).
To try to ensure (b),
the Byzantine coordinator sends a message $\coordval$
with $\neg(r \modulo 2)$ to all non-faulty nodes,
this message is delivered instantly, as a result
all non-faulty processes broadcast an {\sc  aux}
message with value $\neg(r \modulo 2)$.
Then to ensure (d), Byzantine nodes instantly send
{\sc  aux} message with value $\neg(r \modulo 2)$
to a single node.
Furthermore, to ensure (e), Byzantine nodes instantly send
{\sc  aux} message with value $(r \modulo 2)$
to the remaining nodes.
Assuming both $0$ and $1$ entered $\binvalues$
at appropriate times at non-faulty nodes,
termination will be prevented for this round.

The difficulty in ensuring this non-termination scenario
is that the Byzantine nodes do not control the time
that both $1$ and $0$ enter $\binvalues$ of
non-faulty nodes.  If $\neg(r \modulo 2)$
enters too late, a process may broadcast 
$(r \modulo 2)$ as its {\sc  aux} message, and as
a result we may fail with (d).
Otherwise if $(r \modulo 2)$ enters $\binvalues$
too late, all non-faulty processes may terminate
with $n-t$ {\sc  aux} messages with value $\neg(r \modulo 2)$.
Similar timing arguments can be made for other non-terminating
scenarios that use different message patterns.

\subsection{Different experiment configurations}

\begin{figure}
 \includegraphics[scale=0.53]{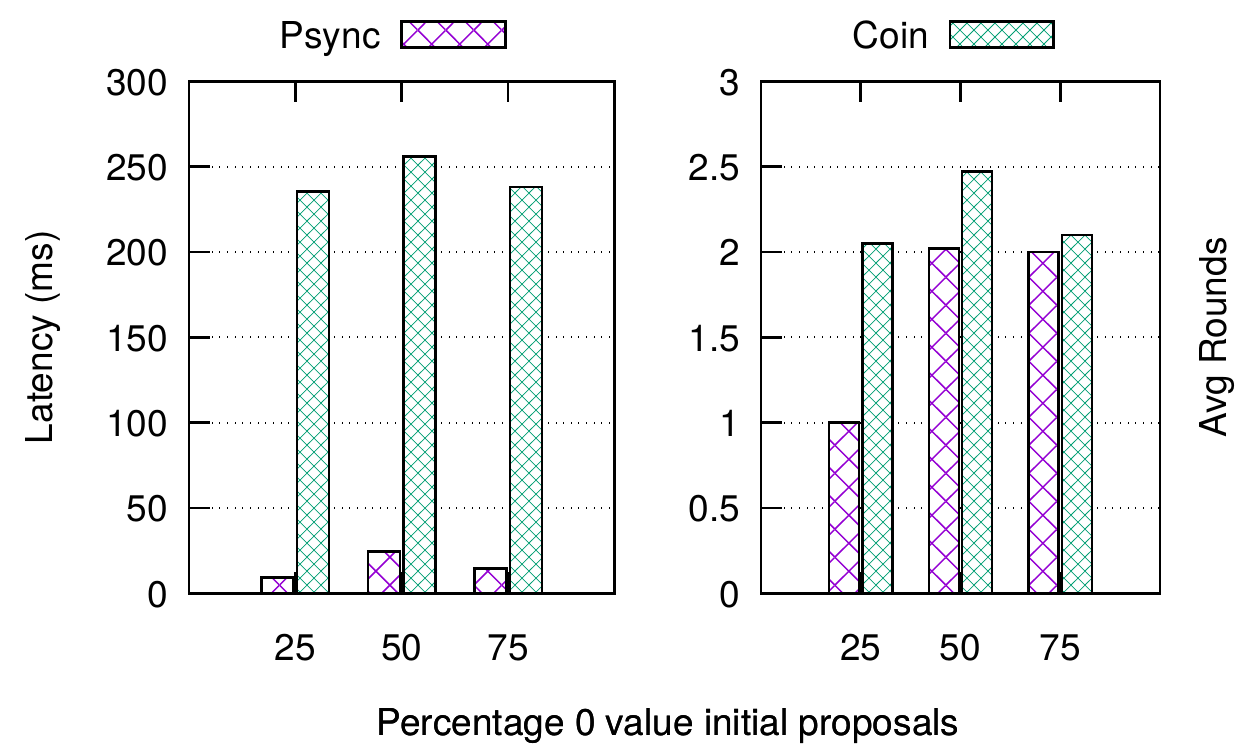}
 \includegraphics[scale=0.53]{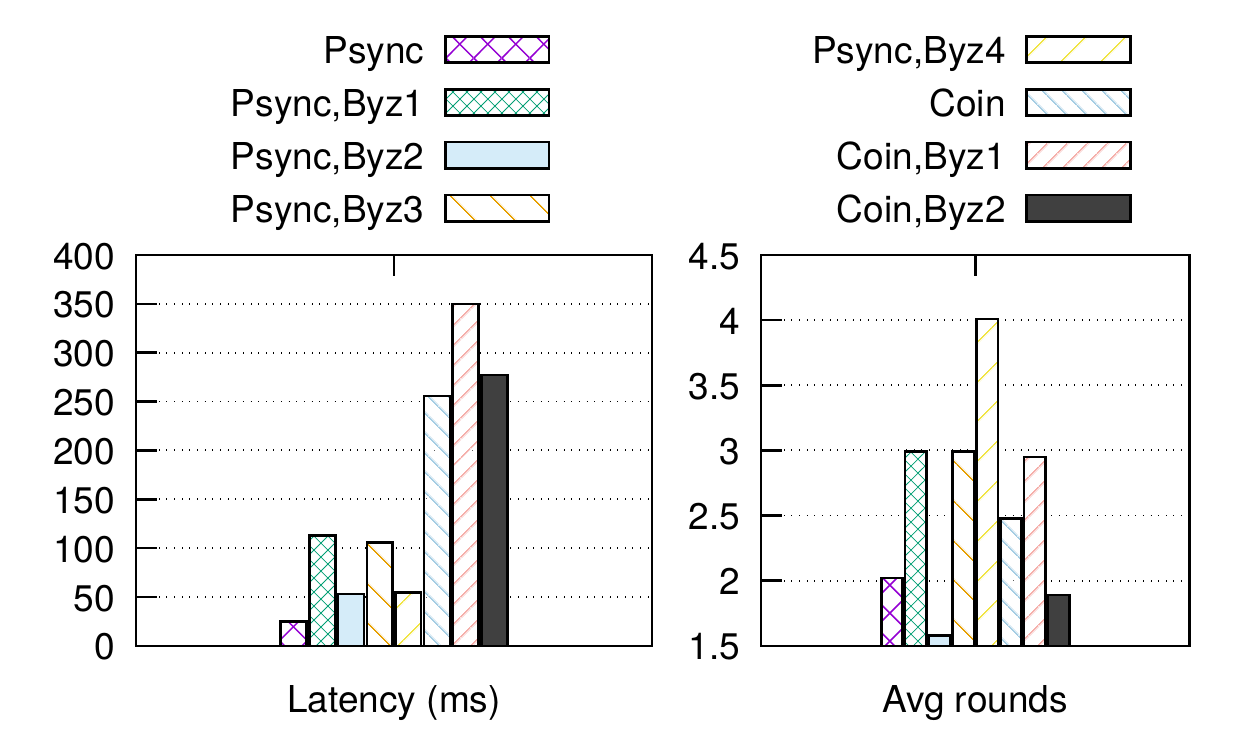}
 \caption{Single datacenter comparison of latency and average number of rounds to terminate of our deterministic binary Byzantine consensus against randomized binary Byzantine consensus: {\bf (Left)} 
 \label{fig:comparison-dc1}}
\end{figure}

Figure~\ref{fig:comparison-dc1} uses the same experimental settings as Figure~\ref{fig:comparison}, except is run with $100$ nodes within a single datacenter.
Here we see a much larger gap in latency between Psync and Coin as the computation of the cryptographic operations of the random coin is much larger
than the network latency.
Note that the latency of both algorithms could be decreased through the use of message authentication codes (MACs) with datagram broadcasts,
but we expect the latency to still be dominated by the crypotgraphic operations of the coin.

Figure~\ref{fig:comparison-dc1} uses the same experimental settings as Figure~\ref{fig:comparison}, except is run with $1$ node
in each of Amazon's $14$ EC2 data centers.
The results are similar to the $5$ datacenter case of Figure~\ref{fig:comparison}, but with higher latency in most cases
due to the increased geo-distribution.


\begin{figure}
\end{figure}


\section{Proofs of safety and liveness of the algorithms}
\label{sec:proofs}

\subsection{Safety proof of the binary Byzantine consensus (Figure~\ref{algo-Bin-Byz-consensus-safe})}
\label{ssec:safe-proof}
The proof is described from a point of view of a non-faulty process $p_i$.
Let $\values_i^r$ denote the
value of the set $\values_i$ which
satisfies the predicate of
line~\ref{BYZ-safe-07} during a round $r$.
Moreover, let us recall
that, given a run, $C$ denotes the set of non-faulty processes in this run.

\begin{lemma}
\label{lemma-BBC-same-value}
Let $t<n/3$. If at the beginning of a round $r$,
all non-faulty processes have the same estimate $v$,
they never change their estimate value thereafter. 
\end{lemma}

\begin{proofL}
Let us assume that all non-faulty processes (which are at least
$n-t>t+1$) have the same estimate $v$ when they start round $r$.
Hence, they all BV-broadcast the same message {\sc est}$[r](v)$ at
line~\ref{BYZ-safe-04}. It follows from the BV-Justification and
BV-Obligation properties that each non-faulty process $p_i$ is such that
$\binvalues_i[r]=\{v\}$ at line~\ref{BYZ-safe-05}, and consequently can
broadcast only {\sc aux}$[r](\{v\})$ at line~\ref{BYZ-safe-06}.
Considering any non-faulty process $p_i$, it then follows from the
predicate of line~\ref{BYZ-safe-07} ($\values_i$ contains only $v$),
the predicate of line~\ref{BYZ-safe-09} ($\values_i$ is a singleton), and
the assignment of line~\ref{BYZ-safe-10}, that $est_i$ keeps the value $v$.
\renewcommand{\toto}{lemma-BBC-same-value}
\end{proofL}

\begin{lemma}
\label{lemma-BBC-same-bin-value-set}
Let $t<n/3$. \big($(p_i, p_j \in C) \wedge 
(\values_i^r=\{v\}) \wedge (\values_j^r=\{w\})$\big)$\Rightarrow (v=w)$.
\end{lemma}

\begin{proofL}
Let $p_i$ be a non-faulty process such that $\values_i^r=\{v\}$.
It follows from line~\ref{BYZ-safe-07} that $p_i$ received the same message
{\sc aux}$[r](\{v\})$ from $(n-t)$ different processes, i.e., from at least
$(n-2t)$ different non-faulty processes. As $n-2t\geq t+1$, this means that
$p_i$  received the message {\sc aux}$[r](\{v\})$ from  a set $Q_i$
including at least $(t+1)$ different non-faulty processes. 

Let  $p_j$ be a non-faulty process such that $\values_j^r=\{w\}$.
Hence,  $p_j$ received {\sc aux}$[r](\{w\})$ from a set $Q_j$
of at least $(n-t)$ different processes.
As $(n-t)+(t+1)>n$, it follows that $Q_i\cap Q_j\neq \emptyset$.
Let $p_k\in Q_i\cap Q_j$. As $p_k\in Q_i$, it is a non-faulty process.
Hence, at  line~\ref{BYZ-safe-06}, $p_k$ sent the same message
{\sc aux}$[r](\{\})$ to $p_i$ and $p_j$, and we consequently have $v=w$. 
\renewcommand{\toto}{lemma-BBC-same-bin-value-set}
\end{proofL}

\begin{lemma}
\label{lemma-BBC-validity}
Let $t<n/3$. The value decided by a non-faulty process was proposed by a
non-faulty process.
\end{lemma}

\begin{proofL}
Let us consider the round $r=1$.
Due to the BV-Justification property of the BV-broadcast of
line~\ref{BYZ-safe-04}, it follows that the sets $\binvalues_i[1]$
contains only values proposed by non-faulty processes. 
Consequently, the non-faulty processes broadcast at line~\ref{BYZ-safe-06}
messages {\sc aux}$[1]()$ containing sets with values proposed only by
non-faulty processes. 
It then follows from the predicate (i) of line~\ref{BYZ-safe-07}
($\values_i^1 \subseteq \binvalues_i[1]$), and the BV-Justification property
of the BV-broadcast abstraction, 
that the set $\values_i^1$ of each non-faulty process contains only values
proposed by  non-faulty processes. Hence, 
the assignment of $est_i$ (be it at line~\ref{BYZ-safe-10}
or~\ref{BYZ-safe-11}) provides it with a value proposed by a non-faulty process.
The same reasoning applies to rounds $r=2$, $r=3$, etc., which concludes the
proof of the lemma. 
\renewcommand{\toto}{lemma-BBC-validity}
\end{proofL}

\begin{lemma}
\label{lemma-BBC-agreement}
Let $t<n/3$. No two  non-faulty processes decide different values.
\end{lemma}

\begin{proofL}
Let $r$ be the first round during which a non-faulty process decides, let
$p_i$ be a non-faulty process that decides in round $r$
(line~\ref{BYZ-safe-10}), and let $v$ be the value it decides.  Hence,
we have $\values_i^r=\{v\}$ where $v=(r \modulo 2)$.
  
If another non-faulty process $p_j$ decides during round $r$, we have
$\values_j^r=\{w\}$, and, due to Lemma~\ref{lemma-BBC-same-bin-value-set},
we have $w=v$. Hence, all non-faulty processes that decide in round $r$,
decide $v$. Moreover, each non-faulty process that decides in round $r$
has previously assigned $v=(r \modulo 2)$  to its local estimate $est_i$.

Let $p_j$ be a non-faulty that does not decide in round $r$.
As  $\values_i^r=\{v\}$, and $p_j$ does not decide in round $r$,
it follows from Lemma~\ref{lemma-BBC-same-bin-value-set}
that  we cannot have  $\values_j^r=\{1-v\}$, and consequently
$\values_j^r=\{0,1\}$. Hence, in round $r$,  $p_j$ executes
line~\ref{BYZ-safe-11}, where it assigns the value $(r \modulo 2)=v$ to
its local estimate $est_j$.

It follows that all non-faulty processes start round $(r+1)$ with the
same local estimate $v=r \modulo 2$. Due to
Lemma~\ref{lemma-BBC-same-value}, they keep this estimate value
forever. Hence, no different value can be decided in a future round by
a non-faulty process that has not decided during round $r$, which
concludes the proof of the lemma.
\renewcommand{\toto}{lemma-BBC-agreement}
\end{proofL}

\begin{lemma}
\label{lemma-BBC-no-stop}
Let the system model be $\BAMP[t<n/3]$. 
No non-faulty process remains blocked forever in a round.
\end{lemma}

\begin{proofL} 
  Let us assume by contradiction that there is
  a first round in which some non-faulty process $p_i$ remains blocked forever. 
  As all non-faulty processes terminate round $(r-1)$, they all start round $r$
  and all invoke  the round $r$ instance of BV-broadcast.
  Due to the BV-Termination property,
  the $\wait()$ statement of  
  line~\ref{BYZ-safe-05} terminates at each non-faulty process. 
  Then, as all non-faulty processes broadcast a message {\sc aux}$[r]()$
  (line~\ref{BYZ-safe-06}), it follows that the  $\wait()$ statement of
  line~\ref{BYZ-safe-07} terminates at each non-faulty process. 
  It follows that there is no first round at which a non-faulty process
  remains blocked forever during round $r$.
\renewcommand{\toto}{lemma-BBC-no-stop}
\end{proofL}

\begin{lemma}
\label{lemma-BBC-single-val-stop}
Let the system model be $\BAMP[t<n/3]$. 
If all non-faulty processes $p_i$ terminate a round $r$
with $\values_i^r=\{v\}$, they all decide by round $(r+1)$.
\end{lemma}

\begin{proofL} 
 If all non-faulty processes are such that  $\values_i^r=\{v\}$,
 and the  round $r$ is such that $v=(r \modulo 2)$,
 it follows from lines~\ref{BYZ-safe-08}-\ref{BYZ-safe-10}
 that (if not yet done) each non-faulty process decides during round $r$.
 
 If $r$ is such that $v\neq(r \modulo 2)$, each non-faulty process
 sets its current estimate to $v$ (line~\ref{BYZ-safe-10}).
 As  during the next round we have $v=((r+1) \modulo 2)$,
 and $\values_i^{r+1}=\binvalues_i[r+1]= \{v\}$ at each non-faulty
 process $p_i$, each  non-faulty process decides during round $(r+1)$.  
\renewcommand{\toto}{lemma-BBC-single-val-stop}
\end{proofL}

\begin{lemma}
\label{lemma-BBC-termination-01}
Let the system model be $\BAMP[t<n/3]$. 
If every non-faulty process $p_i$ terminates a round $r$
with $\values_i^r=\{0,1\}$, then it decides by round $(r+2)$.
\end{lemma}

\begin{proofL}
If every non-faulty processes $p_i$ is such that $\values_i^r=\{0,1\}$,
it executes line~\ref{BYZ-safe-11} during round $r$, and we have
$est_i=(r\mod 2)=v$ when it starts round $(r+1)$.  Due to
Lemma~\ref{lemma-BBC-same-value}, it keeps this estimate forever.  As
all non-faulty processes execute rounds $(r+1)$ and $(r+2)$
(Lemma~\ref{lemma-BBC-no-stop}) and $v=((r+2)\mod 2)$, we have
$\values_i^{r+2}=\{v\}$, at each non-faulty process $p_i$. It follows
that each non-faulty process decides at line~\ref{BYZ-safe-10}.
\renewcommand{\toto}{lemma-BBC-termination-01}
\end{proofL}

\begin{theorem}
\label{theorem:safe-bin-algo}
The algorithm described in Figure~{\em\ref{algo-Bin-Byz-consensus-safe}}
satisfies the safety consensus properties.
\end{theorem}

\begin{proofT}
The proof follows from
Lemma~\ref{lemma-BBC-validity} (BBC-Validity) and
Lemma~\ref{lemma-BBC-agreement} (BBC-Agreement).
\renewcommand{\toto}{theorem:safe-bin-algo}
\end{proofT}


\paragraph*{Decision}
The algorithm described in Figure~\ref{algo-Bin-Byz-consensus-safe}
does not guarantee decision.
This may occur for example when some
non-faulty processes propose $0$, the other non-faulty processes propose
$1$, and the Byzantine processes play double game, each proposing $0$
or $1$ to each non-faulty process, so that it never happens that at the
end of a round all non-faulty processes have either $\values_i=\{0,1\}$,
or they all have $\values_i=\{v\}$ with $v$ either $0$ or $1$. In
other words, if not all non-faulty processes propose the same initial
value, Byzantine processes can make, round after round, some non-faulty
processes have $\values_i=\{0,1\}$, while the rest of non-faulty
processes have $\values_i=\{v\}$, with $v \neq (r \mod 2)$, avoiding
them to decide.\footnote{In the case of the randomized binary
  consensus algorithm of~\cite{MMR15}, the common coin guarantees
  termination with probability 1, because eventually the singleton
  value in $\values_i$ will match the coin.}  

\subsection{Why the safe algorithm does not terminate with $\ES{}$}\label{sec:nontermination}


To circumvent the consensus impossibility~\cite{FLP85} and find a terminating solution, one could be
tempted to consider the ${\BAMP}[t<n/3,\ES{}]$ model and setting a timer, that
increases in each round, by replacing line~\ref{BYZ-safe-05} in Figure~\ref{algo-Bin-Byz-consensus-safe} with a new line called ``New1'' and a modified line~\ref{BYZ-safe-05} called ``M-\ref{BYZ-safe-05}'':
\fbox{
\begin{minipage}[t]{150mm}
\footnotesize
\renewcommand{\baselinestretch}{2.5}
\resetline
\begin{tabbing}
aaaA\=aaA\=aaaA\=aaaA\=aaaaaaaaaA\kill
...\\
(New1)  
        \>\> $timeout_i \leftarrow  timeout_i + 1$; 
           $\set$ $timer_i$  $\sto$  $timeout_i$;\\
           
(M-\ref{BYZ-safe-05}) \>\> $\wait$
 \big($(\binvalues_i[r_i]\neq\emptyset)\wedge (timer_i \mbox{ expired})$\big);\\
 ...
\end{tabbing}
\normalsize
\end{minipage}
}

%
% ...
%
\noindent
In fact, this could seem sufficient to
eventually give
enough time for messages to be delivered.  As we explain below, it
would still be possible for a Byzantine process to wait depending on
the timer of the current round to send a message ${\sf BVAL}({v})$ to
a non-faulty process early enough so that this non-faulty process receives
the message before its local timer expires but too late for this
non-faulty process to rebroadcast it and for other non-faulty processes to
deliver it before their timers expire.

As an example, consider a counter-example of $n=4$ processes among
which $t=1$ process is Byzantine that starts from a round $r$ such that
$r \mod{2} = 1$ with non-faulty processes with estimates 0, 0 and 1. There is an
execution leading to a round $r+1$ where $(r+1)\,\mod{2} = 0$ and non-faulty
processes have estimates 0, 1 and 1. The symmetric of this
counter-example can then be used from round $r+1$ where non-faulty
processes have estimates 0, 1 and 1 to round $r+2$ where non-faulty
processes have estimates 0, 0 and 1. An infinite sequence alternating
this counter-example and its symmetric example illustrates an infinite
execution where no non-faulty process decides.

\begin{figure}[ht!]
  \begin{center}
    \includegraphics[scale=0.6]{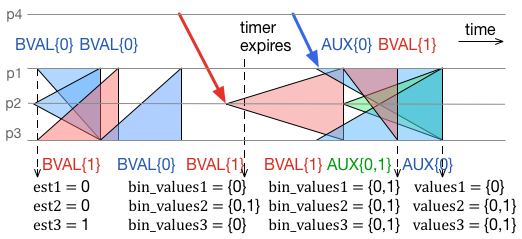} 
	\caption{An execution of $n=4$ processes exchanging broadcast
          messages (represented by triangles between $n-t=3$ non-faulty
          processes) in round $r$ $(r \mod{2} = 1)$ where $est_1 =
          est_2 = 0$ and $est_3 = 1$ leading to a round $r+1$ where
          $est_1 = 0$ and $est_2 = est_3 =
          1$\label{fig:counterexample}}
	\end{center}
\end{figure}

The counter-example is represented as a distributed execution in Figure~\ref{fig:counterexample},
where $p_1$, $p_2$ and $p_3$ are non-faulty processes while $p_4$ is a
Byzantine process, as a distributed execution where time increases
from left to right, where arrows represent messages sent by the
Byzantine process $p_4$ and triangles represent the broadcast messages
among non-faulty processes: the left angle of each triangle indicates the
source of the broadcast while the right edge indicates the
processes 
where messages are delivered. (The
receipt of messages by the Byzantine process $p_4$ are omitted for the
sake of clarity in the presentation.)

The first four triangles represent the BV-broadcast
(Figure~\ref{algo-VB-broadcast}) where $p_1$ and $p_2$ broadcasts
${\sf BVAL}({0})$ while $p_3$ broadcasts ${\sf BVAL}({1})$ according
to their initial estimates.  Once $p_3$ delivers ${\sf BVAL}({0})$
from $t+1=2$ non-faulty processes, it broadcasts the value $0$ that it
never broadcast before as specified in the code of
Figure~\ref{algo-VB-broadcast}.

%
%
%
%
%


During BV-broadcast, all non-faulty processes receive from $2t+1=3$
non-faulty processes.  Now consider that the Byzantine process $p_4$
sends ${\sf BVAL}({1})$ to $p_2$ which makes $p_2$ rebroadcast it as
part of the BV-broadcast because it has now received message ${\sf  BVAL}({1})$
from $2 = t+1$ distinct processes, namely $p_3$ and $p_4$.
We can thus obtain that
$bin\_values_1 = \{0\}$, $bin\_values_2 = \{0,1\}$ and $bin\_values_3 = \{0\}$
at the time non-faulty processes
broadcast their ${\sf AUX}$ messages.  By sending ${\sf AUX}(\{0\})$ to
$p_1$, the Byzantine process $p_4$ allows $p_1$ to choose
$values_1 = \{0\}$ that has received ${\sf AUX}(\{0\})$ from $n-t = 3$ distinct
processes ($p_1$, $p_3$ and $p_4$).
 while the others have to choose
$values_2 = values_3 = \{0,1\}$ as they receive ${\sf AUX}(\{0\})$, ${\sf AUX}(\{0,1\})$,
${\sf AUX}(\{0\})$ from $p_1$, $p_2$ and $p_3$,
respectively.
As $b = r \mod 2 = 1$, it results from line~\ref{BYZ-safe-11} that $p_1$, $p_2$ and $p_3$ have estimates 
0, 1, 1, respectively when starting the round $r+1$. 

Applying the symmetric example would lead to round $r+2$ with the same estimates 0, 0, 1 as in round $r$, indicating the existence of an infinite execution.



\subsection{Proof of Safety and Liveness of the $\ES$-based Binary Byzantine Consensus (Figure~\ref{algo-Bin-Byz-consensus-safe-live})}
\label{sec:safe-live-proof}

The proof consists of two parts: (i) show that the added statements preserve
the consensus safety properties proved for the algorithm
of Figure~\ref{algo-Bin-Byz-consensus-safe}, and (ii) show that
all non-faulty processes eventually decide. 

\begin{lemma}
\label{lemma:keep-safety}
The algorithm described in Figure~{\em\ref{algo-Bin-Byz-consensus-safe-live}}
satisfies the BBC-Validity and BBC-Agreement properties. 
\end{lemma}

\begin{proofL}
The proof consists in showing that the
Lemmas~\ref{lemma-BBC-same-value}, \ref{lemma-BBC-same-bin-value-set},
\ref{lemma-BBC-validity} and \ref{lemma-BBC-agreement} remain correct
when considering the algorithm of
Figure~\ref{algo-Bin-Byz-consensus-safe-live}.  Basically, these
proofs remain correct because, as the new and modified statements
do not assign values to the sets $\binvalues_i[r]$ at the non-faulty processes, 
and no  property of $\binvalues_i$ is related to a timing assumption, 
the set $\binvalues_i[r]$ of a non-faulty process $p_i$
can never contain values proposed by Byzantine
processes only. It follows from this observation that the local
variables $est_i$ and $\values_i$ of any non-faulty process $p_i$
(defined or updated at lines ~M-\ref{BYZ-safe-07}, New5,
\ref{BYZ-safe-10}, or \ref{BYZ-safe-11}) can contain only
values from non-faulty processes.  More specifically we have the following.
\begin{itemize}
\item Lemma~\ref{lemma-BBC-same-value}.
  Let $r$ be the considered round, and 
  $v$ be the current estimate of the non-faulty processes.
  We then have $\binvalues_i[r] = \{v\}$ at line~M-\ref{BYZ-safe-05}
  of every non-faulty process $p_i$. 
\begin{itemize}
\item If the weak round coordinator $p_k$ is non-faulty, we have at
  every non-faulty process  $aux_i=\binvalues_i[r] = \{v\}$. It then follows
  that $\values_i^r=\{v\}$ and the lemma remains true due to 
  lines~\ref{BYZ-safe-09} and~\ref{BYZ-safe-10}.
\item
  If the weak round coordinator $p_k$ is Byzantine and sends possibly different
  values to the non-faulty processes, let us consider a non-faulty process that
  receives the message \coordval$[r](\{1-v\})$.
  As $(1-v)\notin\binvalues_i[r]$, at line  New3,
  $p_i$ executes the ``else'' part where it sets $aux_i$ to $\{v\}$
  (the only value in $\binvalues_i[r]$), and the lemma follows. 
\end{itemize}  
\item Lemma~\ref{lemma-BBC-same-bin-value-set}.  As it
  does not depend on the timers, and is related only to
  the fact that each of the sets $\values_i^r$ and $\values_j^r$ of two
  non-faulty processes are singletons, the proof remains valid.
\item Lemma~\ref{lemma-BBC-validity}.  
  The proof follows from the fact that the sets $\binvalues_i$ of any non-faulty
  process can contain only values proposed by non-faulty processes. 
\item Lemma~\ref{lemma-BBC-agreement}.    
  As it relies only on the set $values_i^r$ of each non-faulty process $p_i$,
  this proof remains correct. 
\end{itemize}
\renewcommand{\toto}{lemma:keep-safety}
\end{proofL}
  
\begin{lemma}
\label{lemma:ensure-decision}
The algorithm described in Figure~{\em\ref{algo-Bin-Byz-consensus-safe-live}}
ensures that every non-faulty process decides. 
\end{lemma}

\begin{proofL}
Let us first observe that, as timers always expire, the ``wait''
statements (modified lines~M-\ref{BYZ-safe-05}
and~M-\ref{BYZ-safe-07}) always terminate, and consequently
Lemma~\ref{lemma-BBC-no-stop} remains true.  The reader can also check
that the proof of Lemma~\ref{lemma-BBC-single-val-stop}
remains valid.

It remains to show that there is eventually a round $r$
at the end of which all non-faulty processes $p_i$  have
the same value $w$ in their set variables ($values_i^r=\{w\}$)
(from which decision follows due to
Lemma~\ref{lemma-BBC-single-val-stop})
The proof shows that, due to (a) the eventual synchrony  assumption, 
(b) the weak round coordinator mechanism, and  (c) the messages \coordval$[~]()$
sent by the weak round coordinators, there is a round $r$
such that $values_i^r=\{w\}$ at each non-faulty process.

Let us consider a time $\tau$ from which (due to Lemma~\ref{lemma:catch-up-first-sync})
the system behaves synchronously (the timeout
values of all non-faulty processes are such that all the messages
exchanged by the non-faulty processes arrive timely).  Let $r$ be the
smallest round number coordinated by a non-faulty process $p_k$ after
$\tau$.  At line New2 of round $r$, $p_k$ broadcasts
\coordval$[r](w)$, being $w$ the first value that enters its set
$\binvalues_k[r]$.  The message \coordval$[r](w)$ is received timely
by all non-faulty processes, that set $aux_i$ to $\{w\}$ in line New3.
Consequently, in line M-\ref{BYZ-safe-06} all non-faulty processes
broadcast {\sc aux}$[r](\{w\})$, and receive in line
M-\ref{BYZ-safe-07} $(n-t)$ {\sc aux}$[r](\{w\})$ messages from
different processes, setting in line New5 $\values_i$ to $\{w\}$.  By
Lemma~\ref{lemma-BBC-single-val-stop}, all non-faulty processes decide
$w$ by round $r+1$, which concludes the proof of the lemma.
\renewcommand{\toto}{lemma:ensure-decision}
\end{proofL} 

\begin{theorem}
\label{theorem:final}
The algorithm described in Figure~{\em\ref{algo-Bin-Byz-consensus-safe-live}}
solves the binary Byzantine consensus in the system model ${\BAMP}[t<n/3,\ES]$.
\end{theorem}

\begin{proofT}
The proof follows directly from Lemma~\ref{lemma:keep-safety}
(BBC-Validity and BBC-Agreement) and Lemma~\ref{lemma:ensure-decision}
(BBC-Termination).
\renewcommand{\toto}{theorem:final}
\end{proofT}



\paragraph*{From asynchrony to synchrony}

In order to guarantee decision, after the eventual synchrony assumption
holds and the timeout value at each non-faulty process is big enough (i.e.,
bigger than the upper bound on message transmission delay), we need that
eventually all non-faulty processes execute rounds synchronously
(as assumed by Lemma \ref{lemma:ensure-decision}). Observe
that, due to initial asynchrony, non-faulty processes can start the
consensus algorithm at different instants. Moreover, due to the
potential participation of Byzantine processes, some non-faulty processes
can advance rounds, without deciding, while other non-faulty processes
are still executing previous rounds.
It is assumed that non-faulty processes may observe
time at different rates and processing time is non-negligible,
but is bounded by some unknown constant.
By using a timeout that grows by $1$ each round
the following proof shows that all processes
eventually reach a round from which they behave synchronously.


%


For the proof we will need to use a
\emph{mini-round} notation and a \emph{catch-up mechanism}.
\begin{itemize}
\item {\bf Mini-round}:
Each round $r$ is split into two mini-rounds, with the first
mini-round representing lines \ref{BYZ-safe-03} to M-\ref{BYZ-safe-05}
and the second representing lines (New3) to \ref{BYZ-safe-12}.
Thus, round 0 is made up of mini-rounds 0 and 1, round 1 is made up of mini-rounds
2 and 3, and so on.
The reason behind splitting the rounds is so that each mini-round
includes a single execution of the timer.

\item {\bf Catch-up mechanism}:
A catch-up mechanism is used to help to the slow
non-faulty processes to catch up to the most advanced non-faulty processes 
(as measured by  their mini-round number).\footnote{Similar mechanisms
are used by PBFT~\cite{CL02}.}
To this end, when a process is in a mini-round $\rho$ and receives
messages corresponding to another mini-round $\rho'$ from
$(t+1)$ different processes (i.e., from at least one non-faulty
process) such that $\rho'>\rho$, the process no longer waits for timers in mini-rounds
$\rho$, .., $(\rho'-1)$.  It still completes these mini-rounds, but
does so without waiting for timers expiration.
\end{itemize}


We assume that each process has a local clock that allows it to measure time units
as integers.
A process uses its local clock to measure the amount of time
it waits for a timeout (where a timeout of $1$ is $1$ time unit).
The notation $t$ with a subscript (for example $t_{\firstzero}$) will be used
to represent a time measurement that is given by the number of time
units that have passed since the algorithm started,  
as measured by an omniscient global observer $G$.
By $\ES$, processes are able to observe time at different rates, but within an
unknown fixed bound.
For simplicity we assume that the fastest non-faulty process observes time at a rate
no faster than observed by the global observer $G$, thus all other processes
observe time at this rate or slower.  The timeouts used in the following proof
are relative to the timeouts of the fastest process.

\paragraph*{Definitions}
The following definitions will be used in the proofs.
\begin{itemize}
\item $\delta$ is a fixed, but unknown bound on message transfer
  delays as ensured by $\ES$ and measured in time units as observed by
  $G$. 
\item $t_{\firstrho}$ is the time, as measured by $G$, at which
  the first non-faulty process $p_{\firstrho}$ reaches mini-round $\rho$
  ($t_{\firstzero}$ is the time at which the first non-faulty process starts
  the consensus).  
\item $t_{\lastrho}$ is the time, as measured by $G$, at which
  the last (i.e. the slowest for that mini-round)
  non-faulty process $p_{\lastrho}$ reaches mini-round $\rho$
  ($t_{\lastzero}$ is the time at which the last non-faulty process starts
  the consensus). 
\item $\theta_{fast}$ (resp. $\theta_{slow}$) is the minimum (resp. maximum)
  amount of time, as observed by
  $G$, for any process to perform the computation of any mini-round
  (an unknown bounded difference between $\theta_{fast}$ and $\theta_{slow}$
  is ensured by $\ES$). 
\item $\gamma_{fast_\rho}$ is the minimum
  amount of time, as observed by $G$, in a mini-round
  $\rho$ that any process waits on line New1 or
  New4 before starting its timer for that
  mini-round. 
\item Mini-round $\rho_\delta$ is the first mini-round where $timeout > \delta$
  at any non-faulty process.
\end{itemize}  

The proof is made up of two lemmas.
Lemma \ref{lemma:catch-up-to-timeout} shows that processes will eventually
reach a point where they remain no more than one mini-round apart.
Lemma \ref{lemma:catch-up-first-sync} builds upon this to show
that the rounds eventually become synchronous.

\begin{lemma}
  \label{lemma:catch-up-to-timeout}
  Consider the algorithm of Figure~{\em\ref{algo-Bin-Byz-consensus-safe-live}} enriched with the
  previous catch-up mechanism.  There is a mini-round $\rho_t$ such
  that in $\rho_t$ and for all following mini-rounds all non-faulty
  processes must wait for at least part of the $timeout$, i.e.,
  they do not receive $t+1$ messages from a mini-round larger than
  $\rho_t$ until after they start waiting for the timeout of mini-round $\rho_t$.
\end{lemma}
    
\begin{proofL}
  Let us consider mini-round $\rho_t$ where
  $\rho_t > \rho_\delta$.  For all non-faulty processes to wait at a
  timeout in a mini-round $\rho_t$, the last non-faulty process to
  arrive at $\rho_t$ must arrive before it receives a
  message from some other non-faulty process that has already started executing a later
  mini-round (note that given $\rho_t > \rho_\delta$, this can only
  occur when the non-faulty processes are no more than $1$ mini-round apart).
  Thus, to satisfy the lemma, a mini-round is needed where the following
  inequality holds at that and all following mini-rounds:
  \begin{equation}\label{inequality:mini-r}
    t_{last_{\rho_t}} < t_{first_{\rho_{t+1}}}.
  \end{equation}
  To find out when this is satisfied first we
  will compute the minimum and maximum times at which non-faulty processes
  can arrive at a mini-round.
  By definition,
  a non-faulty process can spend no less
  time than $(\gamma_{fast_{\rho'}} + \theta_{fast} + timeout_{\rho'})$ in a
  mini-round $\rho'$.  Given that timeouts start with value $0$ in mini-round
  $0$ and grows by $1$ in each mini-round, $timeout$ can be replaced with
  $\rho$ for any mini-round $\rho$ as a lower bound for the fastest process.
  We can then compute the time
  where the first non-faulty process arrives at mini-round $\rho'$
  (where $\rho' > \rho_\delta$) as:
  $$t_{first_{\rho'}} \geq t_{first_{\rho_\delta}} + \left(
  \sum_{x=\rho_\delta}^{\rho'-1} \gamma_{fast_x} + \theta_{fast} + x
  \right).$$
  Notice that from the component 
  $\sum_{x=\rho_\delta}^{\rho'-1} x$ (i.e., the timeout), the value of
  $t_{first_{\rho'}}$ is quadratic in the number of mini-rounds.

  Now consider how long it will take the slowest non-faulty process
  to execute mini-round $\rho'$ when it does not wait at a timeout.
  By definition we know the process will spend no more time than
  $\theta_{slow}$ on computation.
  Thus, the remaining time will be spent waiting until the
  \wait() conditions in the algorithm are satisfied.
  We will now examine how much time a non-faulty process can spend waiting
  during a mini-round
  on either line M-\ref{BYZ-safe-05} or M-\ref{BYZ-safe-07}
  (we only consider these \wait() conditions as they encompass
  the others within a mini-round).
  
  First consider line M-\ref{BYZ-safe-05}. Its condition requires $(\binvalues_i[r_i] \neq \bot)$.
  Given that the process is not waiting at a timeout, it must have
  received $(t + 1)$ messages corresponding to a later mini-round, meaning that some
  non-faulty process has already completed $\rho'$.
  Furthermore, given that this is the slowest non-faulty process, we know that all
  non-faulty processes have already executed the $\BVbroadcast()$ operation on line \ref{BYZ-safe-04}.
  As we can see in Figure \ref{algo-VB-broadcast}, in the $\BVbroadcast()$ operation
  all non-faulty processes will perform at most $2$ broadcast operations.
  Thus, by the BV-Uniformity property, all non-faulty processes will have a value in their
  $\binvalues_i[r_i]$ after at most $2$ message
  delays following the slowest non-faulty processes invocation of the $\BVbroadcast()$.
  As a result, the process takes at most $2 \times \delta + \theta_{slow}$ time to
  execute the mini-round.
  
  
  Now consider line M-\ref{BYZ-safe-07}.
  By the time the slowest non-faulty process has reached this line all non-faulty processes have
  broadcast their {\sc aux} messages, thus the slowest non-faulty process will receive
  these {\sc aux} messages in at most $\delta$ time.
  The process may then need to wait for another message delay
  to satisfy all the conditions of line M-\ref{BYZ-safe-07} in the case
  where a non-faulty process had a value enter its $\binvalues_i[r_i]$
  immediately before broadcasting its {\sc aux} message (recall that the
  $\BVbroadcast()$ may take up to $2$ message delays).
  Thus, as before, the process takes at most $2 \times \delta + \theta_{slow}$ time to
  execute the mini-round.

  We then
  have:
  $$t_{last_{\rho'}} \leq t_{last_{\rho_\delta}} + \left(
  \sum_{x=\rho_\delta}^{\rho'-1} 2 \times \delta + \theta_{slow}
  \right).$$
  Notice that the value of $t_{last_{\rho'}}$ is linear in
  the number of mini-rounds.
  
  Now given $t_{first_{\rho'}}$ is quadratic while $t_{last_{\rho'}}$
  is linear, inequality (\ref{inequality:mini-r}) must eventually be
  satisfied and there will be a mini-round where all non-faulty
  processes wait for at least part of their timeout.

  It will now be shown that for mini-rounds where $timeout > (3 \times
  \delta + \theta_{slow})$, once inequality (\ref{inequality:mini-r}) is true,
  it will remain true for all following mini-rounds.  This will be done by induction.
  Consider $t_{last_{\rho_t}} <
  t_{first_{\rho_{t+1}}}$ is satisfied, let us now show that
  $t_{last_{\rho_t+1}} < t_{first_{\rho_{t+2}}}$ must also be satisfied.
  For this to not hold, the slowest non-faulty process must spend more
  time on mini-round $\rho_t$ than the fastest non-faulty process spends on
  mini-round $(\rho_t +1)$, but this is impossible because once
  the fastest process completes the condition
  on line New1 or New4 and starts its timer,
  $p_{last_{\rho_t}}$ must receive $(t+1)$ messages from mini-round
  $(\rho_t + 1)$ after $\delta$ time.
  Once these messages are received, the process will not
  wait at any timeout, and as we have already seen, the this process will
  take no more than $2 \times \delta +
  \theta_{slow}$ time to complete the mini-round.  Thus, as long as
  $timeout > (3 \times \delta + \theta_{slow})$, which will eventually
  be true given $\ES$ and the growing timeout, process
  $p_{last_{\rho_t}}$ will reach mini-round $(\rho_t + 1)$ before
  $p_{first_{\rho_t+1}}$ reaches mini-round $(\rho_t + 2)$.
  \renewcommand{\toto}{lemma:catch-up-to-timeout}
\end{proofL}

\begin{lemma}
  \label{lemma:catch-up-first-sync}
  Consider the algorithm of Figure~\ref{algo-Bin-Byz-consensus-safe-live} enriched with the
  previous catch-up mechanism.  Eventually the non-faulty processes
  attain a mini-round from which they behave synchronously.
\end{lemma}

\begin{proofL}
  By Lemma \ref{lemma:catch-up-to-timeout} it is known that there
  exists a mini-round $\rho_t$ where at that and all following mini-rounds
  all non-faulty processes wait for at least part of their timeout.
  Additionally, this must happen at some mini-round where
  $timeout > (3 \times \delta + \theta_{slow})$. Consider we are in
  such mini-rounds.  Now for a mini-round to be synchronous,
  all non-faulty processes need to arrive at that mini-round with enough
  time to broadcast their messages to all non-faulty processes before any
  non-faulty process moves onto the next mini-round.
  In the case that the last non-faulty process to arrive at the mini-round is the
  weak coordinator, it may take up to $3$ message delays before its {\coordval}$[r]()$
  message is received by all non-faulty processes (this includes up to $2$ message delays
  until a value enters its $\binvalues[r]$ and an additional
  message delay to broadcast {\coordval}$[r]()$).
  Thus, for a mini-round $\rho_t'$
  to by synchronous where $\rho_t' \geq \rho_t$, the following needs
  to be ensured:
  \begin{equation}\label{inequality:sync-needed}
    t_{last_{\rho_t'}} + (3\times \delta) + \theta_{slow}
                \leq t_{first_{\rho_t'}} + \gamma_{fast_{\rho_t'}} + timeout_{\rho_t'}.
  \end{equation}
  Let us now compute $t_{last_{\rho_t'}}$.
  First, notice that before a non-faulty process starts its timer for a mini-round it
  must wait until the condition on line New1 or
  New4 is satisfied.  
  Also note that by time $(t_{first_{\rho_t'}} + \gamma_{fast_{\rho_t'}} +
  \theta_{fast})$ at least one process has satisfied the condition on
  line New1 or New4 (this is given by
  the definition of $\gamma$).  As a result all processes will receive
  $(t+1)$ messages from mini-round $\rho_t'$ by time
  $(t_{first_{\rho_t'}} + \gamma_{fast_{\rho_t'}} + \theta_{fast} +
  \delta)$.  Now given Lemma \ref{lemma:catch-up-to-timeout} and that
  $(\rho_t' - 1) > \delta$, it is known that that the slowest process
  is no further behind than waiting at the timeout of mini-round
  $(\rho_t' -1)$.  After getting these $(t+1)$ messages from
  mini-round $\rho_t'$ the slow process will then skip the timeout of
  mini-round $(\rho_t' -1)$ and reach the following mini-round in at
  most $2$ additional message delays ($2$ message delays are needed
  for the same reasons given in Lemma \ref{lemma:catch-up-to-timeout}
  to satisfy the condition line M-\ref{BYZ-safe-05} or
  M-\ref{BYZ-safe-07}) plus any processing time.  Thus, the time at
  which the slowest process reaches mini-round $\rho_t'$ is given by:
  $$t_{last_{\rho_t'}} \leq t_{first_{\rho_t'}} +
  \gamma_{fast_{\rho_t'}} + \theta_{fast} + \theta_{slow} + (3 \times
  \delta).$$
  Now plugging this into inequality
  (\ref{inequality:sync-needed}) leads to $timeout_{\rho_t'} \geq (7 \times \delta) + (2 \times
  \theta_{slow}) + \theta_{fast}$ (note that $2 \times \theta_{slow}$
  is included to account for possible processing times in both mini-rounds
  $(\rho_t'-1)$ and $\rho_t'$).  But given that the timeout grows in each
  mini-round and that $\delta$, $\theta_{fast}$, and $\theta_{slow}$
  are bound by $\ES$ there will eventually be a mini-round where this
  holds true.

   Finally, notice that as long as the timeout is this large (i.e. $timeout \geq (7 \times \delta) + (2 \times
  \theta_{slow}) + \theta_{fast}$) and Lemma~\ref{lemma:catch-up-to-timeout} holds then the above argument is
  valid for any mini-round.  Now given that $timeout \geq (7 \times \delta)
  + (2 \times \theta_{slow}) + \theta_{fast}$ is larger than the
  timeout needed for Lemma \ref{lemma:catch-up-to-timeout} to hold for
  every following mini-round, once inequality (\ref{inequality:sync-needed}), i.e. synchrony,
  it true for one mini-round, it will also hold for every following mini-round.
  \renewcommand{\toto}{lemma:catch-up-first-sync}
\end{proofL}

\subsection{Proof of the Blockchain Consensus (Figure~\ref{algo:reduction-multi-to-bin})}


\begin{lemma}
\label{lemma-one-bin-cons-decides-1}
There is at least one binary consensus instance that decides value $1$,
and all non-faulty processes exit the repeat loop. 
\end{lemma}
\noindent
From an operational point of view, this lemma can be re-stated as follows:
there is at least one $\ell\in[1..n]$ such that
at each non-faulty process $p_i$, we eventually have $\bindec_i[\ell]=1$. 

\begin{proofL}
The proof is by contradiction.
Let us assume that, at any non-faulty process $p_i$,
no $\bindec_i[\ell]$, $1\leq \ell \leq n$,
is ever set to $1$ (line~\ref{MV-BYZ-12}).
It follows that no non-faulty process
exits the ``repeat'' loop (lines~\ref{MV-BYZ-02}-\ref{MV-BYZ-04}).
As a non-faulty process $p_j$ RB-broadcasts a valid value, it follows
from the RB-Termination-1 property, that each non-faulty process $p_i$
RB-delivers the  valid proposal of  $p_j$, 
and consequently we eventually have $\proposals_i[j]\neq\bot$
at each non-faulty process $p_i$ (line~\ref{MV-BYZ-11}).


It  follows from the first sub-predicate of line~\ref{MV-BYZ-02}
and the RB-Termination-2 property
that all non-faulty processes $p_i$ invokes $\binpropose(-1)$
on the BBC object  $\BINCONS[j]$ and by line~\ref{MV-BYZ-11},
they all BV-deliver $1$ to round one.
Notice that by using the RB-delivery to trigger
the BV-delivery of $1$ (instead of calling ${\BVbroadcast}$)
the lemma relies on the fact that the properties of ${\RBbroadcast}$
also ensure the properties of ${\BVbroadcast}$.
Namely that RB-Termination-1 ensures
BV-Obligation, RB-Validity ensures BV-Justification,
and RB-Termination-2 ensures BV-Uniformity and BV-Termination.
It follows that the properties of the binary consensus are
maintained.
Hence, from its BBC-Termination, BBC-Agreement,  BBC-Validity,
and Intrusion-tolerance properties (as no non-faulty process
has proposed $0$),
this BBC instance returns the value $1$ to all non-faulty processes,
which exit the ``repeat'' loop.

\renewcommand{\toto}{lemma-one-bin-cons-decides-1}
\end{proofL}

\begin{lemma}
\label{lemma-C-validity}
A decided value is a valid value (i.e., it satisfies the predicate
$\valid()$).
\end{lemma}

\begin{proofL}
Let us first observe that, for a value $\proposals_i[j]$ to be decided
by a process $p_i$, we need to have  $\bindec_i[j]=1$
(lines~\ref{MV-BYZ-08}-\ref{MV-BYZ-10}).

If the value  $1$ is decided by  $\BINCONS[j]$,  $\bindec_i[j]=1$
is eventually true at each non-faulty process $p_i$ 
(line~\ref{MV-BYZ-12}). If follows from
(i) the fact that the value $1$ can only enter the
$\binvalues$ of a BBC instance
after validation at line~\ref{MV-BYZ-11}, and (ii)
the Intrusion-tolerance property of  $\BINCONS[j]$,
that at least one non-faulty process
$p_i$ inserted $1$ into its $binvalues$ on line~\ref{MV-BYZ-12}.
Due to line~\ref{MV-BYZ-11}, it follows that $\proposals_i[j]$ contains
a valid value.
\renewcommand{\toto}{lemma-C-validity}
\end{proofL}

\begin{lemma}
\label{lemma-C-agreement}
No two non-faulty processes decide different values.
\end{lemma}

\begin{proofL}
\sloppy{Let us consider any two non-faulty processes $p_i$ and $p_j$, such that
$p_i$ decides $\proposals_i[k1]$ and $p_j$ decides $\proposals_j[k2]$.
%
It follows from line~\ref{MV-BYZ-08} that
$k1= \mmin \{x \mbox{ such that } \bindec_i[x]=1\}$ and
$k2= \mmin \{x \mbox{ such that } \bindec_j[x]=1\}$.}

\sloppy{On the one hand, it follows from line~\ref{MV-BYZ-07} that
$(\bigwedge_{1\leq x\leq n} \bindec_i[x]\neq \bot)$ and
$(\bigwedge_{1\leq x\leq n} \bindec_j[x]\neq \bot)$, from which we conclude that
both $p_i$ and $p_j$ know the binary value decided by  
each binary consensus instance (line~\ref{MV-BYZ-12}). 
Due to the BBC-Agreement property of each binary consensus instance,
we also have  $\forall x:~ \bindec_i[x]= \bindec_j[x]$.
Let $dec[x]= \bindec_i[x]= \bindec_j[x]$.
It follows then from line~\ref{MV-BYZ-08} that
$k1= k2= \mmin \{x \mbox{ such that } dec[x]=1\}=k$. 
Hence,  $dec[k]=1$.}

On the other hand, it follows from the Intrusion-tolerance property of
$\BINCONS[k]$ that a non-faulty process $p_\ell$
inserted $1$ into its $binvalues$ on line~\ref{MV-BYZ-12}.
As this invocation can be issued only at
line~\ref{MV-BYZ-03}, we conclude (from the predicate of
line~\ref{MV-BYZ-02}) that $\proposals_\ell[k]=v\neq\bot$.  As $p_\ell$
is non-faulty, it follows from the RB-Unicity and RB-Termination-2
properties that all non-faulty processes RB-delivers $v$ from
$p_k$. Hence, we eventually have $\proposals_i[k]=\proposals_j[k]$,
which concludes the proof of the lemma.
\renewcommand{\toto}{lemma-C-agreement}
\end{proofL}

\begin{lemma}
\label{lemma-C-termination}
Every non-faulty process decides a value. 
\end{lemma}

\begin{proofL}
It follows from Lemma~\ref{lemma-one-bin-cons-decides-1} that there is some
 $p_j$ such that   we eventually have   $\bindec_i[j]=1$  at all
non-faulty processes,  and no 
non-faulty process blocks forever at line~\ref{MV-BYZ-04}.
Hence,  all non-faulty processes invoke each binary consensus instance (at
line~\ref{MV-BYZ-03} or line~\ref{MV-BYZ-06}).  Moreover, due to their
BBC-Termination property, each of the $n$ binary consensus instances
returns a result at each non-faulty process (line~\ref{MV-BYZ-12}).  It
follows that no non-faulty process $p_i$ blocks forever at
line~\ref{MV-BYZ-07}.  Finally, as seen in the proof of
Lemma~\ref{lemma-C-agreement}, the predicate of line~\ref{MV-BYZ-09}
is eventually satisfied at each non-faulty process, which concludes the
proof of the lemma.  \renewcommand{\toto}{lemma-C-termination}
\end{proofL}

\begin{theorem}
\label{theorem-multivalued-consensus}
The algorithm described in Figure~{\em\ref{algo:reduction-multi-to-bin}}
implements multivalued Byzantine consensus ({\em VPBC}) in the system model
${\BAMP}[t<n/3,\mbox{\em BBC}]$.
\end{theorem}

\begin{proofT}
Follows from
Lemma~\ref{lemma-C-validity} (VPBC-Validity),
Lemma~\ref{lemma-C-agreement} (VPBC-Agreement), and
Lemma~\ref{lemma-C-termination} (VPBC-Termination).
\renewcommand{\toto}{theorem-multivalued-consensus}
\end{proofT}

\subsection{Complexity}
The proposed reduction has constant time complexity.

\begin{lemma}
\label{lemma-rb-termination1}
When $\RBbroadcast$ is invoked by a non-faulty process, all non-faulty
processes $\RBdeliver$ the value in constant time.
\end{lemma}

\begin{proofL}
Within $\RBbroadcast$, a
process starts by calling $\broadcast$
with a value, which will then be echoed by all
non-faulty processes, resulting in all processes delivering the value in $3$
communication steps.
\renewcommand{\toto}{lemma-rb-termination1}
\end{proofL}

\begin{lemma}
\label{lemma-rb-termination2}
For any process that $\RBdeliver$s a value,
within a constant amount of time following this all non-faulty processes
have $\RBdeliver$d the value.
\end{lemma}

\begin{proofL}
  For a process to $\RBdeliver$ a value it must have received the value from
  $n-t$ processes, thus $t+1$ non-faulty processes have $\broadcast$ this value,
  which means all non-faulty will echo that value and all non-faulty will receive
  the value from $n-t$ processes in at most $2$ communication steps following
  the first $\RBdeliver$y.
\renewcommand{\toto}{lemma-rb-termination2}
\end{proofL}

\begin{theorem}
\label{theorem-consensus-termination}
The reduction presented in figure \ref{algo:reduction-multi-to-bin} is
a constant time reduction.
\end{theorem}

\begin{proofT}
  Let us show this by contradiction.  Assume a non-faulty process does
  not decide in constant time, there are two possibilities how this
  could happen, either: (i) the process waits on line \ref{MV-BYZ-09}
  for more than constant time, or (ii) the process has not invoked
  some instance $\BINCONS[k]$ until after constant time had already
  passed.

  First consider (i).  Here the process is waiting to $\RBdeliver$ a value
  that by lemma \ref{lemma-C-validity} has already been $\RBdeliver$d
  by some non-faulty process, but by lemma \ref{lemma-rb-termination2} we know
  this must happen in constant time.
  
  Now consider (ii).  By lemma \ref{lemma-one-bin-cons-decides-1} we
  know that at least one binary consensus instance decides $1$.  Once
  this happens all non-faulty processes invoke all remaining instances of
  $\BINCONS[k]$ without waiting.  Thus, for (ii) to be true, no
  instance of $\BINCONS[k]$ must have terminated with $1$ in constant
  time.  But given that at least $2t+1$ instances of $\RBbroadcast$
  will be invoked by non-faulty processes, and given lemma
  \ref{lemma-rb-termination1}, all non-faulty processes will invoke all
  instances of $\BINCONS[k]$ in constant time on line \ref{MV-BYZ-03}
  or \ref{MV-BYZ-06}.
  
\renewcommand{\toto}{theorem-consensus-termination}
\end{proofT}


\section{The BV-broadcast all-to-all communication implementation}\label{sec:bvbcast}
Figure~\ref{algo-VB-broadcast} depicts the pseudocode of an existing 
implementation~\cite{MMR15} of the BV-broadcast problem stated in Section~\ref{ssec:bvbcast}.

\begin{figure*}[ht!]
\centering{
\fbox{
\begin{minipage}[t]{150mm}
\footnotesize
\renewcommand{\baselinestretch}{2.5}
\resetline
\begin{tabbing}
aA\=aaA\=aaaaaaaaaaaaA\kill

{\bf opera}\={\bf tion} ${\sf \BVbroadcast}$ {\sc msg}$(v_i)$ {\bf is}\\

\line{BYZ-BV-01} \>  ${\sf broadcast}$  {\sc b\_val}$(v_i)$.\\~\\

{\bf when} {\sc b\_val}$(v)$  {\bf is received}\\

\line{BYZ-BV-02} \>   
   {\bf if} \=
\big({\sc b\_val}$(v)$  received from  $(t+1)$
        different processes and {\sc b\_val}$(v)$ not yet broadcast\big)\\

\line{BYZ-BV-03} \> \>   
{\bf then}  ${\sf broadcast}$ {\sc b\_val}$(v)$    
     {\it {\scriptsize  \hfill \color{gray}{// a process echoes a value only once}}} \\
\line{BYZ-BV-04} \>   
   {\bf if} \=
\big({\sc b\_val}$(v)$  received from $(2t+1)$ different processes\big)\\
\line{BYZ-BV-04} \> \>   
{\bf then} $bin\_values_i \gets bin\_values_i \cup \{v\}.$ $~~~$ 
  {\it {\scriptsize\hfill\color{gray}{// local delivery of a value}}}
%
\end{tabbing}
\normalsize
\end{minipage}
}
\caption{An algorithm implementing BV-broadcast in ${\BAMP}[t<n/3]$
  (from~\cite{MMR14})}
\label{algo-VB-broadcast} 
}
\end{figure*}

\section{Reliable broadcast in Byzantine systems}
\label{sec:reliable-broadcast}

This broadcast abstraction (in short, RB-broadcast) was proposed  by 
G. Bracha~\cite{B87}. It is a one-shot one-to-all communication abstraction, 
which provides processes with  two operations denoted
${\RBbroadcast}()$ and ${\RBdeliver}()$. When $p_i$ invokes  the
operation ${\RBbroadcast}()$ (resp., ${\RBdeliver}()$), 
we say that it ``RB-broadcasts'' a message (resp., ``RB-delivers'' a message).
An RB-broadcast instance, where process $p_x$ is the sender, is defined 
by the following properties.  
\begin{itemize}[noitemsep,nolistsep]
\item RB-Validity. If a non-faulty process RB-delivers a message $m$ 
from 
a non-faulty process $p_x$, then $p_x$ RB-broadcast $m$. 
\item RB-Unicity. A non-faulty process RB-delivers at most one message
  from $p_x$.
\item RB-Termination-1. 
If $p_x$ is non-faulty and RB-broadcasts a message $m$,   
all the non-faulty processes eventually RB-deliver $m$  from $p_x$. 
\item RB-Termination-2. 
If a  non-faulty process RB-delivers a message $m$ from $p_x$ 
(possibly faulty)  then all the non-faulty processes  eventually  RB-deliver 
the same message  $m$ from~$p_x$. 
\end{itemize}

The  RB-Validity property relates the output to the input, while 
RB-Unicity states that there is no message duplication. 
The termination properties state the cases where processes have to 
RB-deliver messages. The second of them is what makes the broadcast reliable. 
It is shown  in~\cite{BT85} that $t<n/3$ is an upper 
bound on $t$ when one  has to implement such an abstraction.

Let us remark that it is possible that a value may be RB-delivered by the
non-faulty process while its sender is actually Byzantine and has not
invoked ${\RBbroadcast}()$. This may occur for example when the
Byzantine sender played at the network level, at which it sent several
messages to different subsets of processes, and the RB-delivery predicate
of the algorithm implementing the RB-broadcast abstraction is eventually
satisfied for one of these messages.  When this occurs, by abuse of language,
we say that the sender invoked RB-broadcast. This is motivated by the
fact that, in this case, a non-faulty process cannot distinguish if the
sender is faulty or not.

The algorithm described in~\cite{B87} implements RB-broadcast in 
${\BAMP}[t<n/3]$. Hence, it is $t$-resilience optimal. This algorithm
requires three communication steps to broadcast an application message.

\remove{

}


\begin{thebibliography}{99}
\footnotesize{

\bibitem{AMN17}
Abraham, I., Malkhi, D., Nayak, K., Ren, L., Spiegelman, A.
Solida: A Blockchain Protocol Based on Reconfigurable Byzantine Consensus.
{\it Proc. 21st International Conference on Principles of Distributed Systems}, pp.~1--19, (2017)

\bibitem{ABB18}
 Androulaki, E., 
                Barger, A.,  Bortnikov, V., 
                Cachin, C.,
                Christidis, K., 
                De Caro, A.,  
                Enyeart, D., 
                Ferris, C.,
                Laventman, G.
                Manevich, Y.,
                Muralidharan, S., 
                Murthy, C.,
                Nguyen, B.,
                Sethi, M.,
                Singh, G., 
                Smith, K.,
                Sorniotti, A.,
                Stathakopoulou, C.,
                Vukolic, M.
                Weed Cocco, S. and
                Yellick, J.
Hyperledger fabric: a distributed operating system for permissioned
               blockchains.
{\it Proc. of the Thirteenth EuroSys Conference, EuroSys 2018, pp.~30:1--30:15, (2018)}


\bibitem{AGZ15}
Aublin P.-L., Guerraoui R., Knezevic N., Quema V., and Vukoli\'{c} M.,
The next 700 BFT protocols.
{\it ACM Transactions on Computer Systems}, 32(4), Article 12, 45 pages (2015)

\bibitem{ABQ13}
Aublin P.-L., Ben Mokhtar, S., Quema V.,
RBFT: Redundant Byzantine Fault Tolerance.
{\it Proc. 33rd Int'l Conference on Distributed Computing Systems} pp.~297--306, (2013)

\bibitem{A03}
Aspnes J.,
Randomized protocols for asynchronous consensus.
{\it Distributed Computing}, 16(2-3):165-175 (2003)


\bibitem{BE03}
Ben-Or M., El-Yaniv R.,
Resilient-optimal interactive consistency in constant time.
{\it Distributed Computing} 16(4): 249-262 (2003)


\bibitem{BKR94}
Ben{-}Or M., Kelmer B., and Rabin T.,
Asynchronous Secure Computations with Optimal Resilience.
{\it Proc. Annual ACM Symposium on Principles} 
pp.~183-192 (1994)

\bibitem{BSA14}
Bessani, A., Sousa, J., Alchieri, E.A.P.,
State Machine Replication for the Masses with BFT-SMART. 
{\it Proc. 44th Annual IEEE/IFIP International Conference on Dependable Systems and Networks}pp.~355-362 (2014)



\bibitem{BS09}
Brief Announcement: A Leader-free Byzantine Consensus Algorithm.
Fatemeh Borran and Andr\'{e} Schiper.
DISC 2009.


\bibitem{B87}
Bracha G.,
Asynchronous Byzantine agreement protocols.
{\it Information \& Computation}, 75(2):130-143 (1987)

\bibitem{BT85}
Bracha G. and Toueg S.,
Asynchronous consensus and broadcast protocols.
{\em Journal of the  ACM}, 32(4):824-840 (1985)

\bibitem{But16}
Buterin V.,
Ethereum: platform review, opportunites and challenges for
private and consortium blockchains (2016)


\bibitem{CGR11}
Cachin C., Guerraoui R., and Rodrigues L.,
{\it Reliable and secure distributed programming}, 
Springer, 367 pages (2011) ISBN 978-3-642-15259-7

\bibitem{CKPS01}
Cachin C., Kursawe K., Petzold F., and Shoup V.,
Secure and Efficient Asynchronous Broadcast Protocols
{\it Proc. 21st Annual International Cryptology Conference (CRYPTO)},
pp.524-541, 2001



\bibitem{CR93}
  Canetti, R.,
  Fast asynchronous Byzantine agreement with optimal resilience.
  \it{STOC 1993}, 42-51 (1993)

\bibitem{CL02}
Castro M. and  Liskov B.,
Practical Byzantine fault tolerance and proactive recovery.
{\it ACM Transactions on  Computer Systems}, 20(4):398-461 (2002)




\bibitem{CT96}
Chandra T. and Toueg S., 
Unreliable failure detectors for reliable distributed systems. 
 {\em Journal of the ACM}, 43(2):225-267 (1996)



\bibitem{CFV06}
Correia M., Ferreira Neves N., and Verissimo P.,
From consensus to atomic broadcast: time-free Byzantine-resistant
protocols without signatures.
{\it The Computer Journal}, 49(1):82-96 (2006)

\bibitem{CGLR18}
Crain, T., Gramoli, V., Larrea, M., Raynal, M.
DBFT: Efficient Byzantine Consensus with a Weak Coordinator and its Application to Consortium Blockchains
\url{http://poseidon.it.usyd.edu.au/~gramoli/web/doc/pubs2/DBFT-TR.pdf}


\bibitem{CWA09}
Clement, A., Wong, E., Alvisi, L., Dahlin, M. and Marchetti, M.
Making Byzantine fault tolerant systems tolerate Byzantine faults.
NSDI (2009).

\bibitem{DDS87}
Dolev D.,  Dwork C. and  Stockmeyer L., 
On the minimal synchronism needed for distributed consensus.
{\it Journal of the ACM}, 34(1):77-97 (1987)


\bibitem{DLS88}
Dwork C., Lynch N., and Stockmeyer L.,
Consensus in the presence of partial synchrony.
{\it Journal of the ACM}, 35(2):288-323 (1988)




\bibitem{FL82}
Fischer M.J. and Lynch N.A.,
A lower bound for the time to assure interactive consistency.
{\em Information Processing Letters}, 14(4):183-186 (1982)


\bibitem{FLP85}
Fischer M.J., Lynch N.A.,  and Paterson M.S.,
Impossibility of distributed consensus with one faulty process.
{\em Journal of the ACM}, 32(2):374-382 (1985)






\bibitem{Gra17}
Vincent Gramoli.
The Red Belly Blockchain.
Invited talk. Facebook, Menlo Park, USA.
\url{http://gramoli.redbellyblockchain.io/web/doc/talks/facebook.pdf}








\bibitem{KMM03}
Kihlstrom K.P., Moser L.E., and Melliar-Smith P.M.,
Byzantine fault detectors for solving consensus.
{\it The Computer Journal}, 46(1):16-35 (2003)

\bibitem{KS16}
King V. and Saia J.,
Byzantine agreement in expected polynomial time.
{\it Journal of the ACM}, 63(2), Article 13, 21 pages (2016)

\bibitem{KAD07}
Kotla R., Alvisi L., Dahlin M., Clement A., and Wong E.L.,
Zyzzyva: speculative Byzantine fault tolerance.
{\it ACM Transactions on  Computer Systems,} 27(4):7:1-7:39 (2009)

\bibitem{Kur00}
Kursawe K.,
Optimistic asynchronous Byzantine agreement. 
Manuscript (2000)


\bibitem{L78}
Lamport L., 
Time, clocks, and the ordering of events in a distributed system.
{\it Communications of the ACM}, 21(7):558-565 (1978)



\bibitem{LSP82}
Lamport L., Shostack R., and Pease M., The Byzantine generals problem.
{\it ACM Transactions on Programming Languages and Systems}, 
4(3)-382-401 (1982)


\bibitem{LVC16} Liu S., Viotti P., Cachin C., Qu{\'{e}}ma V., and
  Vukoli\'{c} M.,
  XFT: practical fault tolerance beyond crashes.
  {\it Proc. 12th USENIX Symposium on Operating Systems Design and
    Implementation (OSDI'16)}, ACM Press, pp.~485-500 (2016)



\bibitem{L96}
Lynch N.A.,
{\it Distributed algorithms}.
Morgan Kaufmann Pub., San Francisco (CA), 872 pages (1996)
ISBN 1-55860-384-4

\bibitem{MA06}
Martin J.-Ph. and Alvisi L.,
Fast Byzantine consensus.
{\it IEEE Transactions on Dependable and Secure Computing}, 3(3):202-215 (2006)


\bibitem{Mic16}
Micali, S.
ALGORAND: The Efficient and Democratic Ledger.
arXiv:1607.01341v7 (2016).

\bibitem{MXC16}
Miller A., Xia Y., Croman K., Shi E., and Song D.,
The Honey Badger of {BFT} Protocols
{\it Proc. of the 2016 {ACM} {SIGSAC} Conference on Computer and Communications Security},
p.31-42 (2016)


\bibitem{MMR14}
Most\'efaoui A., Moumen H., and Raynal M.,
Signature-free Asynchronous Byzantine Consensus with ${T < N/3}$ and ${O(N^2)}$ Messages.
{\it Proc. of the 2014 ACM Symposium on Principles of Distributed Computing}, p.2--9, (2014)

\bibitem{MMR15}
Most\'efaoui A., Moumen H., and Raynal M.,
Signature-free asynchronous binary Byzantine consensus  with $t<n/3$, 
$O(n^2)$   messages, and  $O(1)$ expected time.
{\it Journal of ACM}, 62(4), Article 31, 21 pages (2015)






\bibitem{MR16}
Most\'efaoui A. and  Raynal M.,
Intrusion-tolerant broadcast  and agreement abstractions in the presence 
of  Byzantine processes.
{\it IEEE Transactions on Parallel and Distributed Systems},
27(4):1085-1098 (2016)






\bibitem{N08}
Nakamoto S.,  Bitcoin: a peer-to-peer electronic cash system.
http://www.bitcoin.org (2008)


  
\bibitem{NCV05}
Neves N. F. and Correia M. and Verissimo P.,
Solving vector consensus with a wormhole,
IEEE Transactions on Parallel and Distributed Systems,
16(12):1120-1131 (2005)
  

\bibitem{PSL80}
Pease M., R. Shostak R., and Lamport L.,
Reaching agreement in the presence of faults.
{\it Journal of the ACM}, 27:228-234 (1980)



\bibitem{R10}
Raynal M.,
{\it Communication and agreement abstractions for fault-tolerant 
asynchronous distributed systems}. 
Morgan \& Claypool,  251 pages (2010) ISBN  978-1-60845-293-4 





\bibitem{S90}
Schneider F.B., 
Implementing fault-tolerant services using the state machine approach. 
{\it ACM Computing Surveys}, 22(4):299-319 (1990)

%

\bibitem{ST87}
  Srikanth, T.,
  Simulating Authenticated Broadcasts to Derive Simple Fault-Tolerant Algorithms.
  {\it Distributed Computing}, 2(2): 80-94 (1987)




}
\end{thebibliography}
\end{document}